\documentclass[12pt,preprint]{aastex}

\pdfoutput=1

\newcommand{\githash}{d56324d}\newcommand{\gitdate}{2014-08-28}

\usepackage{color,hyperref}
\definecolor{linkcolor}{rgb}{0,0,0.5}
\hypersetup{colorlinks=true,linkcolor=linkcolor,citecolor=linkcolor,
            filecolor=linkcolor,urlcolor=linkcolor}
\usepackage{url}
\usepackage{amssymb,amsmath}
\usepackage{subfigure}

\newcommand{\project}[1]{\emph{#1}}
\newcommand{\kepler}{\project{Kepler}}
\newcommand{\terra}{\project{TERRA}}

\newcommand{\paper}{\textsl{Article}}

\newcommand{\foreign}[1]{\emph{#1}}
\newcommand{\etal}{\foreign{et\,al.}}
\newcommand{\etc}{\foreign{etc.}}
\newcommand{\True}{\foreign{True}}
\newcommand{\Truth}{\foreign{Truth}}

\newcommand{\figref}[1]{\ref{fig:#1}}
\newcommand{\Fig}[1]{Figure~\figref{#1}}
\newcommand{\fig}[1]{\Fig{#1}}
\newcommand{\figlabel}[1]{\label{fig:#1}}

\newcommand{\Eq}[1]{Equation~(\ref{eq:#1})}
\newcommand{\eq}[1]{\Eq{#1}}
\newcommand{\eqalt}[1]{Equation~\ref{eq:#1}}
\newcommand{\eqlabel}[1]{\label{eq:#1}}
\newcommand{\Sect}[1]{Section~\ref{sect:#1}}
\newcommand{\sect}[1]{\Sect{#1}}
\newcommand{\sectalt}[1]{\ref{sect:#1}}
\newcommand{\App}[1]{Appendix~\ref{sect:#1}}
\newcommand{\app}[1]{\App{#1}}
\newcommand{\sectlabel}[1]{\label{sect:#1}}

\newcommand{\dd}{\ensuremath{\,\mathrm{d}}}
\newcommand{\bvec}[1]{\ensuremath{\boldsymbol{#1}}}
\newcommand{\densityunit}{{\ensuremath{\mathrm{nat}^{-2}}}}

\newcommand{\rate}{\ensuremath{\Gamma}}
\newcommand{\ratepar}{{\ensuremath{\theta}}}
\newcommand{\ratepars}{{\ensuremath{\bvec{\ratepar}}}}
\newcommand{\obs}[1]{\ensuremath{\hat{#1}}}
\newcommand{\radius}{\ensuremath{R}}
\newcommand{\period}{\ensuremath{P}}
\newcommand{\completeness}{{\ensuremath{Q_\mathrm{c}}}}
\newcommand{\transitprob}{{\ensuremath{Q_\mathrm{t}}}}
\newcommand{\data}{{\ensuremath{\bvec{x}}}}
\newcommand{\entry}{{\ensuremath{\bvec{w}}}}

\newcommand{\interim}{{\ensuremath{\bvec{\alpha}}}}

\newcommand{\binarea}{{\ensuremath{\Delta}}}

\newcommand{\mean}{{\ensuremath{\mu}}}
\newcommand{\smooth}{{\ensuremath{\lambda}}}
\newcommand{\smoothpars}{{\ensuremath{\bvec{\smooth}}}}
\newcommand{\cov}{{\ensuremath{\mathrm{K}}}}

\newcommand{\modela}{\emph{Catalog A}}
\newcommand{\modelb}{\emph{Catalog B}}

\newcommand{\gammaearth}{{\ensuremath{\rate_\oplus}}}

\newcommand{\resultsurl}{\url{http://dx.doi.org/10.5281/zenodo.11507}}

\begin{document}

\title{%
    Exoplanet population inference and the abundance \\
    of Earth analogs from noisy, incomplete catalogs
}

\newcommand{\nyu}{2}
\newcommand{\mpia}{3}
\newcommand{\cds}{4}
\newcommand{\princeton}{5}
\newcommand{\berkeley}{6}
\author{%
    Daniel~Foreman-Mackey\altaffilmark{1,\nyu},
    David~W.~Hogg\altaffilmark{\nyu,\mpia,\cds},
    Timothy~D.~Morton\altaffilmark{\princeton}
}
\altaffiltext{1}         {To whom correspondence should be addressed:
                          \url{danfm@nyu.edu}}
\altaffiltext{\nyu}      {Center for Cosmology and Particle Physics,
                          Department of Physics, New York University,
                          4 Washington Place, New York, NY, 10003, USA}
\altaffiltext{\mpia}     {Max-Planck-Institut f\"ur Astronomie,
                          K\"onigstuhl 17, D-69117 Heidelberg, Germany}
\altaffiltext{\cds}      {Center for Data Science,
                          New York University,
                          4 Washington Place, New York, NY, 10003, USA}
\altaffiltext{\princeton}{Department of Astrophysics, Princeton University,
                          Princeton, NJ, 08544, USA}

\begin{abstract}

No true extrasolar Earth analog is known.
Hundreds of planets have been found around Sun-like stars that are either
Earth-sized but on shorter periods, or else on year-long orbits but somewhat
larger.
Under strong assumptions, exoplanet catalogs have been used to make an
extrapolated estimate of the rate at which Sun-like stars host Earth analogs.
These studies are complicated by the fact that every catalog is censored by
non-trivial selection effects and detection efficiencies, and every property
(period, radius, \etc)\ is measured noisily.
Here we present a general hierarchical probabilistic framework for making
justified inferences about the population of exoplanets, taking into account
survey completeness and, for the first time, \emph{observational
uncertainties}.
We are able to make fewer assumptions about the distribution than previous
studies; we only require that the occurrence rate density be a smooth function
of period and radius (employing a Gaussian process).
By applying our method to synthetic catalogs, we demonstrate that it produces
more accurate estimates of the whole population than standard procedures based
on weighting by inverse detection efficiency.
We apply the method to an existing catalog of small planet candidates around G
dwarf stars (Petigura \etal\ 2013).
We confirm a previous result that the radius distribution changes slope near
Earth's radius.
We find that the rate density of Earth analogs is about 0.02 (per star per
natural logarithmic bin in period and radius) with large uncertainty.
This number is much smaller than previous estimates made with the same data
but stronger assumptions.

\end{abstract}


\section{Introduction}
\sectlabel{intro}

NASA's \kepler\ mission has enabled the discovery of thousands of exoplanet
candidates (\citealt{kepler-catalog, burke}).
While many of these candidates have not been confirmed as bona fide planets,
there is evidence that the false positive rate is low (\citealt{morton,
fressin-fp}), enabling conclusions about the population of planets based on
the catalog of candidates.
Many of these planets orbit Sun-like stars (\citealt{petigura}), where the
definition of Sun-like is given in terms of the star's temperature and surface
gravity.
Given these catalogs, it is interesting to ask what we can say about the
population of exoplanets as a function of their physical parameters
(period, radius, \etc).
Observational constraints on the population can inform theories of planet
formation and place probabilistic bounds on the abundance of Earth
analogs\footnote{For our purposes, an ``Earth analog'' is an Earth-sized
exoplanet orbiting a Sun-like star with a year-long period.}.

\citet{petigura} recently published an exoplanet population analysis based on
an independent study of the \kepler\ light curves for 42,557 Sun-like stars.
This study was especially novel because the authors developed their own
planet search pipeline (\terra; \citealt{petigura-a}) and determined the
detection efficiency of their analysis empirically by injecting synthetic
signals into real light curves measured by \kepler.
The occurrence rate function determined by \citet{petigura} agrees
qualitatively with previous studies of small planets orbiting Sun-like stars
(\citealt{dong}).
In particular, both papers describe a ``flattening'' rate function (in
logarithmic radius) for planets around Earth's radius.
Even though no Earth analogs were discovered in their search, \citet{petigura}
used the small candidates that they did find to place an extrapolated
constraint on the frequency of Earth-like exoplanets, assuming a flat
occurrence rate density in logarithmic period.

A very important component of any study of exoplanet populations is the
treatment of detection efficiency.
Speaking qualitatively, in a transit survey, small planets with long periods
are much harder to detect than large planets orbiting close to their star.
This effect is degenerate with any inferences about the rate density and it
can be hard to constrain quantitatively.
In practice, there are three methods for taking this effect into account:
\emph{(a)} making conservative cuts on the candidates and assuming that the
resulting catalog is complete (\citealt{catanzarite, traub, tremaine}),
\emph{(b)} asserting an analytic form for the detection efficiency as a
function of approximate signal-to-noise (\citealt{youdin, howard, dressing,
dong, fressin-fp, morton-swift}), and \emph{(c)} determining the detection
efficiency empirically by injecting synthetic signals into the raw data and
testing recovery (\citealt{pixel, petigura-a, petigura}).

There are two qualitatively different methods that are commonly used to infer
the occurrence rate density from a catalog and a detection efficiency
specification.
The first is an intuitive method that we will refer to as
``inverse-detection-efficiency'' and the second is based on the likelihood
function of the catalog given a parametric rate density.
The inverse-detection-efficiency method involves making a histogram of the
objects in the catalog where each point is weighted by its inverse detection
probability.
This method is very popular in the literature (\citealt{howard, dong,
dressing, swift, petigura}) even though it is not motivated probabilistically.
The alternative likelihood method models the catalog as a Poisson realization
of the \emph{observable} rate density of exoplanets taking the survey
detection efficiencies and transit probabilities into account.
This technique has been used to constrain parametric models---a broken power
law, for example---for the occurrence rate density (\citealt{tabachnik,
youdin, dong}).
In this \paper, we start from the likelihood method but model the rate density
non-parametrically as a piecewise-constant step function.
Using this formulation of the problem, we derive a generalization that takes
observational uncertainties into account.
In \app{inv-det-eff}, we show that the inverse-detection-efficiency method can
be derived as a special case of the likelihood method in the limit of a
smoothly varying completeness function.

In every previous study of exoplanet occurrence rates, the authors have
assumed that the measurement uncertainties are negligible.
This assumption is not justified because these uncertainties---especially on
measurements (like exoplanet radius) that depend on the stellar
parameters---can be large compared to the scales of interest.
In this \paper, we develop a flexible framework for probabilistic inference of
exoplanet occurrence rate density that can be applied to incomplete catalogs
with \emph{non-negligible observational uncertainties}.
Our method takes the form of a hierarchical probabilistic (Bayesian)
inference.
We generalize the method introduced by \citet{hogge} to account for survey
detection efficiencies.
We run tests on simulated datasets---comparing results with the standard
techniques that neglect observational uncertainties---and apply our method to
a real catalog of small planets transiting Sun-like stars
(\citealt{petigura}).

For the purposes of this \paper, we make some strong assumptions, although we
argue that they are weaker than the implicit assumptions in previous
studies.
None of these assumptions is necessary for the validity of our general method
but they do simplify the specific procedures we employ.
We assume that
\begin{itemize}

\item the candidates in the catalog are independent draws from an
inhomogeneous Poisson process set by the censored occurrence rate density,

\item every candidate is a real exoplanet (there are no false positives),

\item the observational uncertainties on the physical parameters are
non-negligible but known (the catalog provides probabilistic constraints on
the parameters),

\item the detection efficiency of the pipeline is known, and

\item the \True\footnote{In this \paper, we use ``\True'' to describe an
observable (for example, the exoplanet occurrence rate density) that would be
trivially measured in the limit of very high signal-to-noise data.
We use ``true'' to describe a simulation quantity with a value exactly known
to us.} occurrence rate density is \emph{smooth}\footnote{We give our
definition of ``smooth'' in more detail below but our model is very flexible
so this is not a strong restriction.}.

\end{itemize}
The first assumption---conditional independence of the candidates---is
reasonable since the dataset that we consider explicitly includes only single
transiting systems (\citealt{petigura}).
The second assumption---neglecting false positives---is also strong and only
weakly justified by estimates of low false positive rates in the \kepler\ data
(\citealt{fressin-fp, morton}).
For this \paper, we will neglect this issue and only comment on the effects
but the prior distributions published by \citet{fressin-fp} could be directly
applied in a generalization of our method.

We must emphasize one very important consequence of our assumptions.
We assume that the catalog of exoplanet candidates is only missing planets
with probabilities given by the empirical detection efficiency.
In detail this must be false; the detection efficiency we use
doesn't take into account the fact that the catalog doesn't include multiple
transiting systems.
A large fraction of the transiting planets discovered by the \kepler\ transit
search pipeline are members of multiple transiting systems (see
\citealt{lissauer}, for example).
Since \citet{petigura} only detected at most one planet per system, their
catalog is actually a list of planet candidates \emph{without a more
detectable companion}.
The global effects of this selection are not trivial and an in-depth
discussion is beyond the scope of this \paper\ but all of the results should
be interpreted with this caveat in mind.

Conditioned on our assumptions and the choices made in the planet detection,
vetting and characterization pipeline (\citealt{petigura-a, petigura}), we
constrain the rate density of small exoplanets orbiting Sun-like stars.
As part of this analysis we also place probabilistic constraints on the rate
density\footnote{In this \paper, we use the word ``rate'' to indicate the
dimensionless expectation value of a Poisson process and the words ``rate
density'' to indicate a quantity that must be integrated over a finite bin in
period and radius to deliver a rate.} of Earth analogs \gammaearth, which we
define as \emph{the expected number of planets per star per natural
logarithmic bin in period and radius, evaluated at the period and radius of
Earth}
\begin{eqnarray}
\gammaearth &=&
\left.\frac{\dd N}{\dd\ln\period\,\dd\ln\radius}\right|
_{\radius=\radius_\oplus,\,\period=\period_\oplus}\quad.
\end{eqnarray}
Since no Earth analogs have been detected, this constraint requires an
extrapolation in both period and radius.
\citet{petigura} performed this extrapolation by assuming that the period
distribution of planets in a small bin in radius is flat, obtaining
$\gammaearth \approx 0.12$.
We relax this assumption and extrapolate only by assuming that the occurrence
rate density is a smooth function of period and radius; we find lower values
for \gammaearth.
We enforce the smoothness constraint by applying a flexible Gaussian process
regularization to the bin heights.

In the next Section, we summarize the likelihood method for exoplanet
population inference and in \sect{hier}, we describe how to include the
effects of observational uncertainties.
The technical term for this procedure is \emph{hierarchical inference} and
while a general discussion of this field is beyond the scope of this \paper,
in \sect{hier}, we present the basic probabilistic question and derive a
computationally tractable inference procedure.
In \sect{model}, we summarize the technique and derive the key equation for
our method: \eq{money}.
We test our method on synthetic catalogs in Sections~\sectalt{valid}
and~\sectalt{extrap}.
In \sect{real}, we use the catalog of planet candidates and the empirically
determined detection efficiency from \citet{petigura} to measure the
occurrence rate density of small planets with long orbital periods.

Sections~\sectalt{likelihood} and~\sectalt{hier} provide a general pedagogical
introduction to the methods used in this \paper.
Readers looking to \emph{implement} a population inference are directed to
\app{inv-det-eff} if measurement uncertainties are negligible or \sect{model}
(especially \eqalt{money}) for problems with non-negligible uncertainties.
Readers interested in our results---the inferred population of exoplanets and
Earth-analogs---can safely skip to \sect{real} and continue to the discussion
in \sect{comparison}.

\section{The likelihood method}
\sectlabel{likelihood}

The first ingredient for any probabilistic inference is a likelihood function;
a description of the probability of observing a specific dataset given a set
of model parameters.
In this particular project, the dataset is a catalog of exoplanet measurements
and the model parameters are the values that set the shape and normalization
of the occurrence rate density.
Throughout this \paper, we use the notation $\rate_\ratepars(\entry)$ for the
occurrence rate density \rate---parameterized by the parameters \ratepars---as
a function of the physical parameters \entry\ (orbital period, planetary
radius, \etc).
In this framework, the occurrence rate density can be ``parametric''---for
example, a power law---or a ``non-parametric'' function---such as a histogram
where the bin heights are the parameters \ratepars.

We'll model the catalog as a draw from the inhomogeneous Poisson process set
by the \emph{observable} rate density $\obs{\rate}_\ratepars$.
This leads to the previously known result (see \citealt{tabachnik,youdin} for
some of the examples from the exoplanet literature)
\begin{eqnarray}\eqlabel{poisson-like}
p(\{\entry_k\}\,|\,\ratepars) &=&
    \exp\left(-\int \obs{\rate}_\ratepars (\entry) \dd\entry\right) \,
    \prod_{k=1}^K \obs{\rate}_\ratepars (\entry_k)\quad.
\end{eqnarray}
In this equation, the integral in the normalization term is the expected
number of observable exoplanets in the sample.

The main thing to note here is that $\obs{\rate}_\ratepars$ is the rate
density of exoplanets that you would expect to observe taking into account the
geometric transit probability and any other detection efficiencies.
In practice, we can model the observable rate density as
\begin{eqnarray}\eqlabel{obs-rate}
\obs{\rate}_\ratepars(\entry) &=&
    \completeness(\entry)\,\rate_\ratepars(\entry)
\end{eqnarray}
where $\completeness(\entry)$ is the detection efficiency (including transit
probability) at \entry\ and $\rate_\ratepars(\entry)$ is the object that we
want to infer: the \True\ occurrence rate density.
We haven't yet discussed any specific functional form for
$\rate_\ratepars(\entry)$ and all of this derivation is equally applicable
whether we model the rate density as, for example, a broken power law or a
histogram.

The observed rate density \obs{\rate} is a quantitative description of the
rate density at which planets appear in the \citet{petigura} catalog; it is
not a description of the \True\ rate density of exoplanets.
Inasmuch as the detection efficiency $\completeness(\entry)$ is calculated
correctly, the function $\rate_\ratepars(\entry)$ will represent the \True\
rate density of exoplanets, at least where there is support in the data.
In practice, an estimate of the detection efficiency will not include every
decision or effect in the pipeline and as this function becomes more accurate,
our inferences about the \True\ rate density $\rate_\ratepars(\entry)$ will be
less biased.

For the results in this \paper, we will assume that the completeness function
$\completeness(\entry)$ is known empirically on a grid in period and radius
but that is not a requirement for the validity of this method.
Instead, we could use a functional form for the completeness and even infer
its parameters along with the parameters of the rate density.

Finally, we model the rate density as a piecewise constant step function
\begin{eqnarray}\eqlabel{rate-model}
\rate_\ratepars (\entry) &=& \left\{\begin{array}{ll}
\exp(\ratepar_1) & \entry \in \binarea_1,\\
\exp(\ratepar_2) & \entry \in \binarea_2,\\
\cdots \\
\exp(\ratepar_J) & \entry \in \binarea_J,\\
0 & \mathrm{otherwise}
\end{array}\right.
\end{eqnarray}
where the parameters $\ratepar_j$ are the log step heights and the bins
$\binarea_j$ are fixed \foreign{a priori}.
In \app{inv-det-eff}, we use this parameterization and derive the analytic
maximum likelihood solution for the step heights.
This result is similar to and just as simple as the
inverse-detection-efficiency method and it is guaranteed to provide a lower
variance estimate of the rate density than the standard procedure.

One major benefit of expressing the problem of occurrence rate inference
probabilistically is that it can now be formally extended to include the
effects of observational uncertainties.

\section{A brief introduction to hierarchical inference}
\sectlabel{hier}

The general question that we are trying to answer in this \paper\ is:
\emph{what constraints can we put on the occurrence rate density of exoplanets
given all the light curves measured by \kepler?}
In the case of negligible measurement uncertainties, this is equivalent to
optimizing \eq{poisson-like} but when this approximation is no longer valid,
we must instead compute the \emph{marginalized likelihood}
\begin{eqnarray}\eqlabel{crazylike}
p(\{\data_k\}\,|\,\ratepars) &=&
    \int p(\{\data_k\}\,|\,\{\entry_k\})
    \,p(\{\entry_k\}\,|\,\ratepars)
    \dd\{\entry_k\}
\end{eqnarray}
where $\{\data_k\}$ is the set of all light curves, one light curve $\data_k$
per target $k$, \ratepars\ is the vector of parameters describing the
population occurrence rate density $\rate_\ratepars(\entry)$ and $\entry_k$ is
the vector of physical parameters describing the planetary system (orbital
periods, radius ratios, stellar radius, \etc) around target $k$.
In this equation, our only assumption is that the datasets depend on the
rate density of exoplanets only through the catalog $\{\entry_k\}$.
In our case, this assumption qualitatively means that the signals found in the
light curves depend only on the actual properties of the planet and star, and
not on the distributions from which they are drawn.
It is worth emphasizing that---as we will discuss further below---the catalog
only provides \emph{probabilistic constraints} on $\{\entry_k\}$; not perfect
delta-function measurements.

In other words, we treat the catalog as being a dimensionality reduction of
the raw data with all the relevant information retained.
In the context of \kepler, the catalog reduces the set of downloaded time
series (approximately 70,000 data points for the typical \kepler\ target) to
probabilistic constraints on a handful of physical parameters---\entry\ from
above---like the orbital period and planetary radius.
If we take this set of parameters $\{\entry_k\}$ as \emph{sufficient
statistics} of the data then we can, in theory, compute \eq{crazylike}---up to
an unimportant constant---without ever looking at the raw data again!
This is important because the high-dimensional integral in \eq{crazylike}
won't generally have an analytic solution and each evaluation of the
per-object likelihood $p(\data_k\,|\,\entry_k)$ is expensive, making numerical
methods intractable.

Instead, we will reuse the hard work that went into building the catalog.
We must first notice that each entry in a catalog is a representation of the
posterior probability
\begin{eqnarray}\eqlabel{crazypost}
p(\entry_k\,|\,\data_k,\,\interim) &=&
\frac{p(\data_k\,|\,\entry_k)\,p(\entry_k\,|\,\interim)}
     {p(\data_k\,|\,\interim)}
\end{eqnarray}
of the parameters $\entry_k$ conditioned on the observations of that object
$\data_k$.
The notation \interim\ is a reminder that the catalog was produced under a
specific choice of a---probably ``uninformative''---\emph{interim prior}
$p(\entry_k\,|\,\interim)$.
This prior was chosen by the author of the catalog and it is different from
the likelihood $p(\entry_k\,|\,\ratepars)$ from \eq{poisson-like}.

Now, we can use these posterior measurements to simplify \eq{crazylike} to a
form that can, in many common cases, be evaluated efficiently.
To find this result, multiply the integrand in \eq{crazylike} by
\begin{eqnarray}
\frac{p(\{\entry_k\}\,|\,\{\data_k\},\,\interim)}
     {p(\{\entry_k\}\,|\,\{\data_k\},\,\interim)} &=&
\prod_{k=1}^K
\frac{p(\entry_k\,|\,\data_k,\,\interim)}
     {p(\entry_k\,|\,\data_k,\,\interim)}
\end{eqnarray}
and use \eq{crazypost} to find
\begin{eqnarray}\eqlabel{simplemarglike}
\frac{p(\{\data_k\}\,|\,\ratepars)}{p(\{\data_k\}\,|\,\interim)} &=&
    \int
    \frac{p(\{\entry_k\}\,|\,\ratepars)}{p(\{\entry_k\}\,|\,\interim)}\,
    p(\{\entry_k\}\,|\,\{\data_k\},\,\interim)
    \dd\{\entry_k\} \quad.
\end{eqnarray}
The data only enter this equation through the posterior constraints provided
by the catalog $\{\entry_k\}$!
For our purposes, this is the \emph{definition} of hierarchical inference.

The constraints in \eq{crazypost} can always be---and often are---propagated
as a list of $N$ samples $\{\entry_k\}^{(n)}$ from the posterior
\begin{eqnarray}\eqlabel{samples}
\{\entry_k\}^{(n)} &\sim& p(\{\entry_k\}\,|\,\{\data_k\},\,\interim) \quad.
\end{eqnarray}
We can use these samples and the Monte Carlo integral approximation to
estimate the marginalized likelihood from \eq{simplemarglike}---up to an
irrelevant constant---as
\begin{eqnarray}\eqlabel{importance}
p(\{\data_k\}\,|\,\ratepars) &\approx&
    \frac{Z_\interim}{N} \sum_{n=1}^N
    \frac{p(\{\entry_k\}^{(n)}\,|\,\ratepars)}
         {p(\{\entry_k\}^{(n)}\,|\,\interim)}
\end{eqnarray}
where the constant $Z_\interim = p(\{\data_k\}\,|\,\interim)$ is not a
function of the parameters \ratepars.
This is very efficient to compute as long as an evaluation of
$p(\{\entry_k\}\,|\,\ratepars)$ is not expensive.
That being said, \eq{importance} could be a high variance estimator of
\eq{simplemarglike}, depending on the number of independent samples $N$ and
the initial choice of $p(\{\entry_k\}\,|\,\interim)$.
Additionally, the support of $p(\{\entry_k\}\,|\,\ratepars)$ in $\{\entry_k\}$
space is restricted to be narrower than that of
$p(\{\entry_k\}\,|\,\interim)$.
Besides this caveat, in the limit of infinite samples, the approximation in
\eq{importance} becomes exact.
\Eq{importance} is the \emph{importance sampling approximation} to the
integral in \eq{simplemarglike} where the trial density is the posterior
probability for the catalog measurements.

A very simple example is the familiar procedure of making a histogram.
If you model the function $p(\{\entry_k\}\,|\,\ratepars)$ as a piecewise
constant rate density---where the step heights are the parameters---and if the
uncertainties on the catalog are negligible compared to the bin widths then
the maximum marginalized likelihood solution for \ratepars\ is a histogram of
the catalog entries.
The case of non-negligible uncertainties is described by \citet{hogge} using a
method similar to the one discussed here.

\section{Model generalities}
\sectlabel{model}

Now, we can substitute \eq{poisson-like} into \eq{simplemarglike} and apply
the importance sampling approximation (\eqalt{importance}) to derive the
following expression for the marginalized likelihood
\begin{eqnarray}\eqlabel{money}
\frac{p(\{\data_k\}\,|\,\ratepars)}{p(\{\data_k\}\,|\,\interim)} &\approx&
    \exp\left(-\int \obs{\rate}_\ratepars (\entry) \dd\entry\right) \,
    \prod_{k=1}^K
    \frac{1}{N_k} \sum_{n=1}^{N_k}
    \frac{\obs{\rate}_\ratepars (\entry_k^{(n)})}
         {p(\entry_k^{(n)}\,|\,\interim)}
\end{eqnarray}
where the values $\{{\entry_k}^{(n)}\}$ are samples drawn from the posterior
probability
\begin{eqnarray}
{\entry_k}^{(n)} &\sim& p(\entry_k\,|\,\data_k,\,\interim)
\end{eqnarray}
as described in the previous section.
\Eq{money} is the \emph{money equation} for our method.
It lets us efficiently compute the \emph{marginalized likelihood of the entire
set of light curves for a particular occurrence rate density}.

In this equation, we're making the further assumption that the catalog treated
the objects independently.
This is a somewhat subtle point if we were to consider targets with more than
one transiting planet---a point that we will return to below---but for the
considerations of the dataset considered here, it is a justified
simplification.

For the remainder of this \paper, we model the rate density as a
two-dimensional histogram with fixed logarithmic bins in period and radius.
When we include observational uncertainties---using \eq{money}---the maximum
likelihood result is no longer analytic.
Therefore, if we want to compute the ``best-fit'' rate density, we can use a
standard non-linear optimization algorithm.

In the regions of parameter space that we tend to care about, the completeness
is low and there are only a few observations with large uncertainties.
In this case, we're especially interested in probabilistic constraints on the
occurrence rate density; not just the best-fit model.
To do this, we must apply a prior $p(\ratepars)$ on the rate density
parameters and generate samples from the posterior probability
\begin{eqnarray}\eqlabel{posterior}
p(\ratepars\,|\,\{\data_k\}) &\propto&
    p(\ratepars)\,p(\{\data_k\}\,|\,\ratepars)
\end{eqnarray}
using Markov chain Monte Carlo (MCMC).

There is a lot of flexibility in the choice of functional form of
$p(\ratepars)$.
In the well-sampled parts of parameter space there are a lot of
detected planets and the choice of prior makes little difference, but in the
regions that we care about, the detection efficiency is low and applying a
prior that captures our beliefs about the rate density is necessary.
This will be especially important when we extrapolate the rate density
function to the location of Earth---in \sect{extrap}---where no exoplanets
have been found.
Therefore, instead of using an uninformative prior, we want to use a prior
that encourages the occurrence rate density to be ``smooth'' but it should be
flexible enough to capture structure that is supported by the data.
To achieve this, we model the logarithmic step heights as being drawn from a
Gaussian process (\citealt{gp, gibson-gp, dfm-gp}).
This model encodes our prior belief that, on the grid scale that we consider,
the rate density should be smooth but it is otherwise very flexible about the
form of the function.

Mathematically, the Gaussian process density is
\begin{eqnarray}
p(\ratepars) &=& p(\ratepars\,|\,\mean,\,\smooth) \nonumber\\
&=& \mathcal{N} \left[\ratepars;\,\mean\,\bvec{1},\,
\cov(\{\binarea_j\},\,\smoothpars)\right]
\eqlabel{gp}
\end{eqnarray}
where $\mathcal{N}(\cdot;\,\mean\,\bvec{1},\,\cov)$ is a $J$-dimensional
Gaussian\footnote{$J$ is the total number of bins.} with a constant mean
\mean\ and covariance matrix \cov\ that depends on the bin centers
$\{\binarea_j\}$ and a set of hyperparameters $\smoothpars = (\smooth_0,\,
\smooth_\period,\,\smooth_\radius)$.
The covariance function that we use is an anisotropic, axis-aligned
exponential-squared kernel so elements of the matrix are
\begin{eqnarray}
\cov_{ij} &=& \smooth_0\,\exp\left(-\frac{1}{2}\,
    [\binarea_i-\binarea_j]^\mathrm{T}\,\Sigma^{-1}\,[\binarea_i-\binarea_j]
\right)
\end{eqnarray}
where $\Sigma^{-1}$ is the diagonal matrix
\begin{eqnarray}
\Sigma^{-1} &=& \left(\begin{array}{cc}
1/\smooth_\period^2 & 0 \\
0 & 1/\smooth_\radius^2
\end{array}\right) \quad.
\end{eqnarray}

The Gaussian process model for the step heights given in \eq{gp} is very
flexible but the results will depend on the values of the hyperparameters
\mean\ and \smoothpars.
Therefore, instead of fixing these parameters to specific values, we add
another level to our hierarchical probabilistic model and marginalize over
this choice.
In other words, we apply priors---uniform in the logarithm---on \mean\ and
\smoothpars, and sample from the joint posterior
\begin{eqnarray} \eqlabel{joint-posterior}
p(\ratepars,\,\mean,\,\smoothpars\,|\,\{\data_k\}) &\propto&
    p(\mean,\,\smoothpars)\,p(\ratepars\,|\,\mean,\,\smoothpars)\,
    p(\{\data_k\}\,|\,\ratepars) \quad.
\end{eqnarray}
Strictly speaking, in this model, $p(\ratepars\,|\,\mean,\,\smoothpars)$ can't
really be called a ``prior'' anymore and the constraints on the step heights
are no longer independent.

There is an efficient algorithm called elliptical slice sampling (ESS;
\citealt{ess, ess-hyper}) for sampling the step heights \ratepars\ from the
density in \eq{joint-posterior}.
In practice, for problems with this specific structure, ESS outperforms more
traditional MCMC methods commonly employed in astrophysics
(e.g.,~\citealt{emcee}).
Our implementation is adapted from Jo Bovy's BSD licensed ESS
code\footnote{\url{https://github.com/jobovy/bovy_mcmc/blob/master/bovy_mcmc/%
elliptical_slice.py}}.
To simultaneously marginalize over the hyperparameter choice, we use the
Metropolis--Hastings update from Algorithm 1 in \citet{ess-hyper}.
We tune the Metropolis--Hastings proposal by hand until we get an acceptance
fraction of $\sim0.2-0.4$ for the hyperparameters.

For all the results below, we run a Markov chain with $10^6$ steps for the
heights and update the hyperparameters every 10 steps.
We only keep the final $2 \times 10^5$ steps and discard the earlier samples
as burn-in.
By estimating the empirical integrated autocorrelation time of the chain
(\citealt{goodman}), we find that the resulting chain has $\gtrsim4000$
independent posterior samples.
These samples provide an approximation to the marginalized probability
distribution for \ratepars.

\section{Data and completeness function}
\sectlabel{data}

Using an independent exoplanet search and characterization pipeline,
\citet{petigura} published a catalog of 603 planet candidates orbiting stars
in their ``Sun-like'' sample of \kepler\ targets.
For each candidate, \citet{petigura} used Markov chain Monte Carlo to sample
the posterior probability density for the radius ratio, transit duration, and
impact parameter assuming uninformative uniform priors.
They then incorporated the uncertainties in the stellar radius and published
constraints on the physical radii of their candidates.
Given this data reduction and since we don't have access to the individual
posterior constraints on radius ratio and stellar radius, we can't directly
compute the importance weights $p(\{\entry_k\}\,|\,\interim)$ needed for
\eq{importance}.
For the rest of this \paper, we'll make the simplifying assumption that these
weights are constant in log-period and log-radius but the results don't seem
to be sensitive to this specific choice.

\citet{petigura} did not publish or share posterior samples of their
measurements of the physical parameter (\eqalt{samples}).
They did publish a list of periods, radii and radius uncertainties based on
their analysis.
Assuming that there is no measurement uncertainty on the period measurement
and that the radius posterior is Gaussian in linear radius (with a standard
deviation given by the published uncertainty), we draw 512 samples for
$\entry_k$ and use these as an approximation to the posterior probability
function.

A huge benefit of this dataset is that Erik Petigura and collaborators
published a rigorous analysis of the empirical end-to-end completeness of
their transit search pipeline.
Instead of choosing a functional form for the detection efficiency of the
pipeline as a function of the parameters of interest, \citet{petigura}
injected synthetic signals of known period and radius into the raw aperture
photometry and determined the empirical recovery after the full analysis.

We use all the injected samples from \citet{petigura} to compute the mean
(marginalized) detection efficiency in bins of $\ln\period$ and $\ln\radius$.
In each bin, this efficiency is simply the fraction of recovered injections.
For the purposes of this \paper, we neglect the counting uncertainties
introduced by the finite number of samples used to estimate the completeness.
The largest injected signal had a radius of $16\,R_\oplus$ but, because of the
measurement uncertainties on the radii, we need to model the distribution at
larger radii.
To do this, we approximate the survey completeness for $\radius>16\,R_\oplus$
as 1.

Given our domain knowledge of how detection efficiency depends on the physical
parameters, the intuitive choice would be to measure the survey completeness
in radius ratio or signal-to-noise instead of period and radius.
It is also likely that a change of coordinates would yield a higher precision
result.
That being said, it is still correct to measure the completeness in period and
radius, and there are a few practical reasons for our choice.
The main argument is that since the radius uncertainties are dominated by
uncertainties in the stellar parameters, it is not possible to use the
published catalog (\citealt{petigura}) to compute constraints on radius
ratios.
In the future, this problem would be solved by publishing a representation of
\emph{the full posterior density function for each object in the catalog}.
In this case, the most useful data product would be \emph{posterior samples
for each target's radius ratio and stellar radius}.

The detection efficiency also depends on the geometric transit probability
$R_\star/a$.
Since we are modeling the distribution in the period--radius plane, we need to
compute the transit probability marginalized over stellar radius and mass.
This marginalized distribution scales only with the period of the orbit as
$\propto \period^{-2/3}$.
In theory, this marginalization should be over the \True\ distribution of
these parameters in the selected stellar catalog but we'll approximate it by
the empirical distribution; a reasonable simplification given the size of the
dataset.
At a period of 10 days\footnote{This period is chosen arbitrarily because the
power law only needs to be normalized at one point.}, the median transit
probability in the selected sample of stars is $5.061\%$ so we model the
transit probability\footnote{We are using the letter $Q$ to indicate
probabilities since we are already using $P$ to mean period.} as a function of
period as
\begin{eqnarray}\eqlabel{transitprob}
\transitprob (\period) &=&
    0.05061\,\left[\frac{\period}{10\,\mathrm{days}}\right]^{-2/3} \quad.
\end{eqnarray}
This expression is clearly only valid for $\period \gtrsim 1.4\,\mathrm{days}$
but the dataset that we are using (\citealt{petigura}) explicitly only
includes periods longer than five days so this is not a problem.
We're using the \emph{median} transit probability (instead of the mean)
because it is a more robust estimator in the presence of outliers but in our
experiments, the results do not seem to be very sensitive to this choice.

Implicit in the expression for the transit probability in \eq{transitprob} is
the assumption that all of the planets are on circular orbits.
Recently, \citet{kipping} demonstrated that when eccentric orbits are
included, our given value is an underestimate by about 10\%.
This effect will propagate directly to our inferred rate densities.
Even though the degeneracy is not exact---due to our choice of priors on the
rate density parameters---it is not a bad approximation to assume that it is
and scale the results down by your preferred factor.
The right thing to do would be to marginalize over this effect directly during
inference but that exercise is beyond the scope of the current \paper.
To complicate matters, the detection probability of a transit is also a
non-trivial function of the duration.
To account for this effect, so non-circular orbits should also be injected
when measuring the survey completeness.

\section{Validation using synthetic catalogs}
\sectlabel{valid}

In order to get a feeling for the constraints provided by our method and to
explore any biases introduced by ignoring the observational uncertainties, we
start by ``observing'' two synthetic catalogs from qualitatively different
known occurrence rate density functions.
For each of these simulations, we take the completeness function computed by
\citet{petigura} as given.
In general, \eq{poisson-like} can be sampled using a procedure called thinning
(\citealt{poisson}) but for our purposes, we'll simply consider a piecewise
constant rate density evaluated on a fine grid in log-period and log-radius.
For this discrete function, the generative procedure is simple;
\begin{enumerate}
{\item loop over each grid cell $i$,}
{\item draw Poisson random integer $K_i\sim\mathrm{Poissson}(\obs{\rate}_i)$
with the observable rate density in the cell, and}
{\item distribute $K_i$ catalog entries in the cell randomly.}
\end{enumerate}
We then choose fractional observational uncertainties on the radii from the
\citet{petigura} catalog and apply them to the true catalog as Gaussian noise.

We generate synthetic catalogs from two qualitatively different rate density
functions.
Both distributions are generated by a separable model
\begin{eqnarray}
\rate_\ratepars (\ln\period,\,\ln\radius) &=&
    \rate_\ratepars^{(\period)}(\ln\period)\,
    \rate_\ratepars^{(\radius)}(\ln\radius)
\end{eqnarray}
but fit using the full general model.
The first catalog---\modela---is generated assuming a smooth occurrence
surface where both distributions are broken power laws.
The second---\modelb---is designed to be exactly the distribution inferred by
\citet{petigura} in the range that they considered and then smoothly
extrapolated outside that range.
The catalogs generated from these two models are shown in \fig{smooth-results}
and \fig{simulation-results}, respectively and the data are available
online\footnote{\resultsurl}.

For each catalog, we directly apply both the inverse-detection-efficiency
procedure as implemented by \citealt{petigura}\footnote{Our implementation
reproduces their results when applied to the published catalog.} and our
probabilistic method, marginalizing over the hyperparameters of the Gaussian
process regularization.
\Fig{smooth-results} and \fig{simulation-results} show the results of this
analysis in both cases.
In particular, the side panels compare the marginalized occurrence rate
density in period and radius to the true functions that were used to
generate the catalogs.
\Fig{smooth-results} shows that even if the \True\ rate density is a smooth
function, the density inferred by the inverse-detection-efficiency method can
appear to have sharp features.
In this first example---where the true distribution is well described by our
Gaussian process model---the probabilistic inference of the occurrence rate
density is both more precise and accurate.

In the second example, the true rate density includes a sharp feature chosen
to reproduce the result published by \citet{petigura}.
In this case, \fig{simulation-results} shows that the probabilistic
constraints on the rate density are less precise but more accurate than
results using the inverse-detection-efficiency method.
This effect is most apparent in the parts of parameter space where the
detection efficiency is low---long period and small radius.

When applied to either simulated catalog, the inverse-detection-efficiency
method gives a high-variance estimate of the true occurrence rate density.
One effect of this variance is that the inferred distribution will appear to
have more small-scale structure than the true underlying distribution.

\section{Extrapolation to Earth}
\sectlabel{extrap}

As well as inferring the occurrence distribution of exoplanets, this dataset
can also be used to constrain the rate density of Earth analogs.
Explicitly, we constrain the occurrence rate density of exoplanets orbiting
``Sun-like'' stars\footnote{In this \paper, we adopt the \citet{petigura}
sample of G-stars as our definition of ``Sun-like''.}, evaluated at the
location of Earth:
\begin{eqnarray}\eqlabel{gammaearth}
\gammaearth &=& \rate (\ln\period_\oplus,\,\ln\radius_\oplus) \\
&=&
\left.\frac{\dd N}{\dd\ln\period\,\dd\ln\radius}\right|
_{\radius=\radius_\oplus,\,\period=\period_\oplus}\quad.
\end{eqnarray}
That is, \gammaearth\ is the rate density of exoplanets around a Sun-like
star (expected number of planets per star per natural logarithm of period per
natural logarithm of radius), evaluated at the period and radius of Earth.

In \eq{gammaearth}, we use the symbol $\Gamma$ instead of the more commonly
used $\eta$ since we define ``Earth analog'' in terms of measurable quantities
with no mention of habitability or composition.
This might seem unsatisfying but the composition of an exoplanet is
notoriously difficult to measure even with large uncertainty and any
definition of habitability is still extremely subjective.
With this in mind, we stick to the observable definition for this \paper.

Since no Earth analogs have been found, any constraints on this density must
be extrapolated from the existing observations.
This is generally done by assuming a functional form for the occurrence rate
density, constraining it using the observed candidates and extrapolating.
All published extrapolations are based on rigid models of the occurrence rate
density (for example, a power law) fit to the catalog and evaluated at the
location of Earth (\citealt{catanzarite, traub}).
\citet{petigura} used their catalog of planet candidates to constrain the rate
of Earth analogs in a specific period--radius bin assuming an extremely rigid
model: \emph{flat in logarithmic period}.
These results are all sensitive to the choice of extrapolation function and
the specific definition of ``Earth analog''.

We weaken the assumptions necessary for extrapolation by only assuming that
the distribution is smooth using the Gaussian process regularization described
in \sect{model}.
Under this model, the occurrence rate density at periods and radii where no
objects have been detected will be constrained---with large uncertainty---by
the heights of nearby bins.
Therefore, even though there are no candidates that qualify as Earth analogs,
we simply fit our model of the occurrence rate density in a large enough
region of parameter space (including Earth) and compute the posterior
constraints on \gammaearth.
This works because the Gaussian process regularization actually captures our
prior beliefs about the shape of the rate density function.
This model---and any other extrapolation---will, of course, break down if
there is an unmeasured sharp feature in the occurrence rate density near the
location of Earth but our method is the most conservative extrapolation
technique published to date.

For comparison, we also implemented and applied the extrapolation technique
applied by \citet{petigura}.
Their method assumes that, for small planets ($1 \le \radius/\radius_\oplus <
2$) on long periods ($\period > 50\,\mathrm{days}$), the occurrence rate
density is a flat function of logarithmic period or, equivalently, the
cumulative rate is linear.
\citet{petigura} used the candidates in their catalog to estimate the slope of
the empirical cumulative period distribution and used that function to
extrapolate.
Instead of defining \gammaearth\ differentially, as we did in \eq{gammaearth},
\citet{petigura} constrained the integral of the rate density over a box in
period and radius ($1 \le \radius/\radius_\oplus < 2$ and $200 \le
\period/\mathrm{day} < 400$).
Since their model implicitly assumes a constant rate density across the bin,
the differential rate is just their number divided by the bin volume.
This rate density (rate divided by bin volume) is what is shown as a
comparison to our results in the figures.

Figures~\figref{smooth-rate} and~\figref{simulation-rate} compare our results
and the results of the \citet{petigura} extrapolation procedure when applied
to the synthetic catalogs.
Since these catalogs were simulated from a known population model, we know the
true value of \gammaearth\ and it is indicated in the figures with a vertical
gray line.
In both cases, our method returns a less precise but more accurate result for
the rate density and the error bars given by the functional extrapolation
are overly optimistic.
One major effect that leads to this bias is that the period distribution is
not flat.
Restricting the result to only include uniform models is equivalent to
applying an extremely informative prior that doesn't have enough freedom to
capture the complexity of the problem.
As a result, the posterior constraints on \gammaearth\ are dominated by this
prior choice and the resulting uncertainties are much smaller than they should
be.

\section{Results from real data}
\sectlabel{real}

Having developed this probabilistic framework for exoplanet population
inferences and demonstrating that it produces reasonable results when applied
to simulated datasets, we now turn to real data.
As described in \sect{data}, we will use the catalog of small exoplanet
candidates orbiting Sun-like stars published by \citet{petigura}.
This is a great test case because those authors empirically measured the
detection efficiency of their pipeline as a function of the parameters of
interest.

We directly applied our method to the \citet{petigura} sample and generated
MCMC samples from the posterior probability for the occurrence rate density
step heights, marginalizing over the hyperparameters of the Gaussian process
model.
The resulting MCMC chain is available online\footnote{\resultsurl}.

\Fig{real-results} shows posterior samples from the inferred occurrence rate
density as a function of period and radius conditioned on the catalog.
The marginalized distributions are qualitatively consistent with the
occurrence rate density measured using the inverse-detection-efficiency
method with larger uncertainties.

The period distribution integrated over various radius ranges is shown in
\fig{period}.
In agreement with \citet{dong}, we find that the period distribution of large
planets ($R > 8\,R_\oplus$) is inconsistent with the distribution of smaller
planets.
The rate density of large planets appears to monotonically increase as a
function of log period while the distribution for small planets seems to turn
over at a relatively short period (around 50 days) and decrease for longer
periods.

The equivalent results for the radius distribution are shown in
Figures~\figref{radius} and~\figref{linear-radius}.
\Fig{radius} shows the log-radius occurrence rate density integrated over
various logarithmic bins in period.
The distributions in each period bin are qualitatively consistent; the
rate density is dominated by small planets (around two Earth radii) with
potential ``features'' near $\radius\sim3\radius_\oplus$ and $\radius\sim
10\radius_\oplus$.
These features appear in every period bin.
They were also detected---using a completely different dataset and
technique---by \citet{dong} and a similar result is visible in the occurrence
rate determined by \citet[][their Figure 7]{fressin-fp} at low
signal-to-noise.
\Fig{linear-radius} shows the same result but presented as a function of
linear radius.
In these coordinates, the rate density in a single bin is no longer
uniform; instead, scales as inverse radius.

Our constraint on the rate density of Earth analogs (as defined in
\sect{extrap}) is in tension---even though our result has large fractional
uncertainty---with the result from \citet{petigura}.
This is shown in \fig{real-rate} where we compare the marginalized posterior
probability function for \gammaearth\ to the published value and uncertainty.
Quantitatively, we find that the rate density of Earth analogs is
\begin{eqnarray}\eqlabel{ge-result}
\gammaearth &=& 0.019^{+0.019}_{-0.010}~\densityunit
\end{eqnarray}
where the ``\densityunit'' indicates that this quantity is a rate density, per
natural logarithmic period per natural logarithmic radius.
Converted to these units, \citet{petigura} measured
$0.119_{-0.035}^{+0.046}~\densityunit$ for the same quantity (indicated as the
vertical lines in \fig{real-rate}).
This rate density is \emph{exactly} what Petigura's extrapolation model
predicts but, for comparison, we can also integrate our inferred rate density
over their choice of ``Earth-like'' bin ($200 \le \period/\mathrm{day} < 400$
and $1 \le \radius/\radius_\oplus < 2$) to find a \emph{rate} of
Earth analogs.
The published rate is $0.057_{-0.017}^{+0.022}$ (\citealt{petigura}) and our
posterior constraint is
\begin{eqnarray}
\int_{\period=200\,\mathrm{day}}^{400\,\mathrm{day}}
\int_{\radius=1\,\radius_\oplus}^{2\,\radius_\oplus}
\rate_\ratepars (\ln\period,\,\ln\radius)
\dd[\ln\radius]
\dd[\ln\period]
&=&
0.019_{-0.008}^{+0.010}
\quad.
\end{eqnarray}

Although they are mainly nuisance parameters, we also obtain posterior
constraints on the hyperparameters \mean\ and \smoothpars.
In particular, the constraints on the length scales in $\ln \period$ and $\ln
\radius$ are $\smooth_\period = 3.65 \pm 1.03$ and $\smooth_\radius = 0.65 \pm
0.12$ respectively.
Both of these scales are larger than a bin in their respective dimension.
For completeness we also find the following constraints on the other
hyperparameters
\begin{eqnarray}
\mean = 5.44 \pm 1.56 &\quad\mathrm{and}\quad&
\ln\smooth_0 = 1.68 \pm 0.72 \quad.
\end{eqnarray}
The MCMC chains used to compute these values is available
online\footnote{\resultsurl}.

\section{Comparison with previous work}
\sectlabel{comparison}

Our inferred rate density of Earth analogs (\eqalt{ge-result}) is not
consistent with previously published results.
In particular, our result is completely inconsistent with the earlier result
based on \emph{exactly the same dataset} (\citealt{petigura}).
This inconsistency is due to the different assumptions made and the detailed
cause merits some investigation.
The two key differences between our analysis and previous work are \emph{(a)}
the form of the extrapolation function, and \emph{(b)} the presence of
measurement uncertainties on the planet radii.

To make their estimate of \gammaearth, \citet{petigura} asserted a flat
distribution in logarithmic period for small planets.
Our results suggest that the data \emph{do not support} this assumption (see
\fig{period}).
We find that the data require a \emph{decreasing} period distribution in the
relevant range.
A similar result was also found by \citet{dong} and it is apparent in Figure 2
of \citet{petigura}.

To test the significance of the choice of extrapolation function, we relax the
assumption of a uniform period distribution and allow the distribution to be
linear in the same range ($R=1-2\,R_\oplus$ and $P=50-400\mathrm{d}$).
Under this model, the likelihood of the catalog of planets in this range can
be calculated using \eq{poisson-like}.
We apply uniform priors in the physically allowed range of slopes and
intercepts for this distribution and estimate the posterior probability for
the extrapolated rate using MCMC (\citealt{emcee}).
This results give a much more uncertain and substantially lower estimate for
the rate of Earth analogs
\begin{eqnarray}
\Gamma_\oplus &=& 0.072^{+0.088}_{-0.047} \quad.
\end{eqnarray}
With the large error bars, this result is consistent with both results (see
\fig{comparison} where this value is labeled ``linear extrapolation'') but it
does not fully account for the discrepancy.

To examine the effects of measurement uncertainties, we repeat our analysis
with the error bars on the radii artificially set to zero, keeping everything
else the same.
This analysis (labeled ``uncertainties ignored'' in \fig{comparison}) gives
the result
\begin{eqnarray}
\Gamma_\oplus &=& 0.040^{+0.031}_{-0.019} \quad.
\end{eqnarray}
This result is relatively more precise and higher than our final result and
consistent with the value obtained with linear extrapolation.
This confirms the hypothesis that the discrepancy between our result and the
previously published values is the combined result of both of our key
generalizations.

For comparison, we have also included the value of \gammaearth\ implied by
\citet[][their Table 2]{dong}.
This result is based on a power law fit to the period distribution of small
planets ($R=1-2\,R_\oplus$) on long periods ($P=10-250\,\mathrm{d}$) in a
different catalog (\citealt{kepler-catalog}) with a parametric completeness
model.
There are a few factors to consider when comparing to this to our analysis.
Firstly, while \citet{dong} fit a power law in log period, this is still a
very restrictive model when considering this large range of periods.
A broken power law might be more applicable.
Furthermore, their analysis did not incorporate the effects of measurement
uncertainties.
Finally, unlike the \citet{petigura}, the \citet{kepler-catalog} catalog used
by \citet{dong} includes multiple transiting systems.
As mentioned previously, the effect of this selection is hard to determine
without further investigation but it should, intuitively, cause any inference
based on the \citet{petigura} sample to be an underestimate of the \True\
rate.

\section{Discussion}

We have developed a hierarchical probabilistic framework for inferring the
population of exoplanets based on noisy incomplete catalogs.
This method incorporates systematic treatment of observational uncertainties
and detection efficiency.
One major benefit of this framework is that it provides the best possible
probabilistic measurements of the population under the assumptions listed in
\sect{intro} and repeated below.
After demonstrating the validity of our method on two qualitatively different
synthetic exoplanet catalogs, we run our inference on a published catalog of
small exoplanet candidates orbiting Sun-like stars (\citealt{petigura}) to
determine the occurrence rate density these planets as a function of period
and radius.
We extrapolate this measurement to the location of Earth and constrain the
rate density of Earth analogs with large error bars.
In order to perform this extrapolation, we don't assume a specific functional
form for the rate density.
Instead, we only assume that it is a smooth function of logarithmic period and
radius.

The occurrence rate density function that we infer is qualitatively consistent
with previously published results using different inference techniques
(\citealt{dong, fressin-fp, petigura}).
In particular, we find (see \fig{radius}) previously recorded features in the
radius distribution around $\radius\sim 3\,\radius_\oplus$ and $\radius\sim
10\,\radius_\oplus$, although not at high signal-to-noise.
We find that the period distributions for planets in different radius bins are
different, in qualitative agreement with previous results (\citealt{dong}).
\Fig{period} shows that larger planets tend to be on longer periods than
smaller planets.


Our extrapolation of the rate density to the location of Earth is more general
and conservative than any previously published method.
We find a rate density of Earth analogs that is inconsistent with the result
published by \citet{petigura}.
This discrepancy can be attributed to both the rigidity of the assumptions
about the period distribution and the effects of non-negligible measurement
uncertainties.
Our extrapolation is also less confident than previous measurements.
Again, this difference is due to the fact that we allow a much more flexible
extrapolation function.
This is another illustration that, against the standard data analysis
folklore, the correct use of flexible models is \emph{conservative}.

In contrast to previous work, we don't define ``Earth analog'' in terms of
habitability or composition.
Instead, we advocate for a definition in terms of more directly observable
quantities (in this case, period and radius).
Furthermore, we define \gammaearth\ as a rate density (per star per
logarithmic period per logarithmic radius) so that its value doesn't depend on
choices about the ``Earth-like'' bin.

In our analysis we make a few simplifying assumptions.
Every assumption has an effect on the results and could be relaxed as an
extension of this project.
For completeness, we list and discuss the effects of our assumptions below.
\begin{itemize}

\item {\bf Conditional independence}\quad
We assume that every object in the catalog is a conditionally independent
draw from the observable occurrence rate density.
This is a bad assumption when applying this method to a different catalog
where multiple transiting systems are included.
In practice, the best first step towards relaxing this assumption is probably
to follow \citet{tremaine} and assume that the mutual inclination distribution
is the only source of conditional dependence between planets.
For this \paper, the assumption of conditional independence is justified
because the dataset explicitly includes only systems with a single transiting
exoplanet.

\item {\bf False positives}\quad
In our inferences, we assume that all of the candidates in the catalog are
\True\ exoplanets.
The rate of false positives in the \kepler\ catalog has been shown to be low
but not negligible (\citealt{morton, fressin-fp}).
Since some of the objects in the catalog are probably false positives, our
inferences about the occurrence rate density are biased high but without
explicitly including a model of false positives, it's hard to say in detail
what effect this would have on the distributions.
In an extension of this work, we could incorporate the effects of false
positives by switching to a mixture model (see \citealt{hogg-line}, for
example) where each object is modeled as a mixture of \True\ exoplanet and
false positive.
In this mixture model, the false positives would be represented using prior
distributions similar to those used by \citet{morton-fp} or
\citet{fressin-fp}.

\item {\bf Known observational uncertainties}\quad
To apply the importance sampling approximation to the published catalog, we
assume that the measurement uncertainties are known and, in this case,
Gaussian.
The assumption of normally distributed uncertainties could be relaxed given
a sampling representation of the posterior probability function for the
physical parameters (period, radius, \etc).
There is recent evidence that the stellar radii of \kepler\ targets might, on
average, be underestimated (\citealt{bastien}), introducing another source of
noise.
It is possible to relax the noise model and include effects like this but
inference would be substantially more computationally expensive.

\item {\bf Given empirical detection efficiency}\quad
\citet{petigura} determined the end-to-end detection efficiency of their
planet detection pipeline as a function of \True\ period and radius by
injecting synthetic signals into real light curves and testing recovery.
We used these simulations as an exact representation of the detection
efficiency of the catalog but there are several missing components.
The biggest effect is probably the fact that this formulation doesn't include
the selection of only the \emph{most detectable signal} in each light curve.
This bias will be largest in the parts of parameter space where the baseline
detection efficiency is lowest: at long periods and small radius.
As a result, our inferences (and the results from \citealt{petigura}) about
the occurrence rate of small planets on long periods is probably
\emph{underestimated} relative to \Truth.
In detail there is another limitation due to the fact that the stellar
parameters are only known noisily and the transit light curve only constrains
the radius ratio.
This means that the marginalized detection efficiency should be measured as a
function of radius ratio and the interpretation in terms of \True\ radius is
only approximately correct.
Given the size of the dataset and the number of injection simulations, this
effect should be small.

\item {\bf Smooth rate function}\quad
Throughout our analysis, we make the prior assumption that the occurrence rate
density is a smooth function of logarithmic period and radius.
This model is useful because it allows us to make probabilistically justified
inferences about the exoplanet population in regions of parameter space with
low detection efficiency.
The assumption that the rate density should be smooth is intuitive but there
is no theoretical indication that it must be true at all scales.
That being said, the Gaussian process regularization that we use to enforce
smoothness is flexible enough to capture substantial departures from smooth if
they were supported by the data.

\end{itemize}
Our assumptions are severe but we believe that this is the most conservative
population inference method currently on the market.

Under the assumptions that we have made here, our inference of the occurrence
rate density of exoplanets places a probabilistic constraint on the number of
transiting Earth analogs in the existing \kepler\ dataset.
If we adopt the definition of ``Earth-like'' from \citet[][$200 \le
\period/\mathrm{day} < 400$ and $1 \le \radius/\radius_\oplus < 2$]{petigura},
and integrate the product inferred rate density function and the geometric
transit probability (\eqalt{transitprob}) over this bin, we find that the
expected number of Earth-like exoplanets transiting the stars in the sample of
Sun-like stars chosen by \citet{petigura} is
\begin{eqnarray}\eqlabel{ntransit}
N_{\oplus,\,\mathrm{transiting}} &=& 10.6_{-4.5}^{+5.9}
\end{eqnarray}
where the uncertainties are only on the expectation value and don't include
the Poisson sampling variance.
This is an exciting result because it means that, if we can improve the
sensitivity of exoplanet search pipelines to small planets orbiting on long
periods, then we should find some Earth analogs in the existing data.
Furthermore, because of the treatment of multiple transiting systems in the
catalog, the \True\ expected number of transiting Earth-like exoplanets
orbiting Sun-like stars is almost certainly larger than the values in
\eq{ntransit}!

Some of the caveats on the results in this paper are due to assumptions made
for computational simplicity but a much more robust study would be possible
given a complete representation of the posterior probability function for the
physical parameters in the catalog.
The use of MCMC to fit models to observations is becoming standard practice in
astronomy and the results in many catalogs (including \citealt{petigura}) are
given as statistics computed on posterior samplings.
For the sake of hierarchical inferences like the method presented here, it
would be very useful if the authors of upcoming catalogs also published
samples from these distributions \emph{along with the value of their prior
function evaluated at each sample}.
In this spirit, we have released the results of this paper as posterior
samplings\footnote{\resultsurl} for the occurrence rate density function.

All of the code used in this project is available from
\url{http://github.com/dfm/exopop} under the MIT open-source software license.
This code (plus some dependencies) can be run to re-generate all of the
figures and results in this \paper; this version of the paper was generated
with git commit \texttt{\githash} (\gitdate).

\acknowledgments
We would like to thank Erik Petigura (Berkeley) for freely sharing his data
and code.
It is a pleasure to thank
Ruth Angus (Oxford),
Tom Barclay (NASA Ames),
Jo Bovy (IAS),
Eric Ford (PSU),
David Kipping (CfA),
Ben Montet (Caltech/Harvard), and
Scott Tremaine (IAS)
for helpful contributions to the ideas and code presented here.
We would also like to acknowledge the anonymous referee and the Scientific
Editor, Eric Feigelson, for suggestions that
substantially improved the paper.
This project was partially supported by the NSF (grant AST-0908357), NASA
(grant NNX08AJ48G), and the Moore--Sloan Data Science Environment at NYU.
This research builds on ideas generated at a three-week workshop supported by
NSF Grant DMS-1127914 to the Statistical and Applied Mathematical Sciences
Institute.
This research made use of the NASA \project{Astrophysics Data System}.

\appendix

\section*{APPENDIX}

\section{Inverse-detection-efficiency}
\sectlabel{inv-det-eff}

One huge benefit of the inverse-detection-efficiency procedure is its
simplicity.
Therefore, it's worth noting that there is a probabilistically justified
procedure that will always provide less biased results while being only
marginally more complicated.

The standard procedure involves making a weighted histogram of the catalog
entries where the weight for object $\entry_k$ is $1/\completeness(\entry_k)$.
This makes intuitive sense but it does not have a clear probabilistic
justification or interpretation.
As we will show below, the maximum likelihood result involves weighting the
points by the inverse of the \emph{integral} of the completeness function over
the bin area.

To motivate this derivation, let's start by considering the following
pathological example: a single bin where the completeness sharply drops from
one to zero halfway across the bin.
If we observe $K$ objects in this bin, we would have observed about $2K$
objects in a complete sample.
If we apply the inverse-detection-efficiency procedure to this dataset, each
sample will get unit weight because they are all found in the part of the bin
where the completeness is one.
Therefore, we would \emph{underestimate} the true rate in the bin by half.
It's clear in this specific case that giving the points a weight of two would
give a better solution and we'll derive the general result below.

If we model the occurrence rate density as a histogram with $J$ fixed bin
volumes $\binarea_j$ (\eqalt{rate-model}) then \eq{poisson-like} becomes
\begin{eqnarray}
\ln p(\{\entry_k\}\,|\,\ratepars) &=&
    \sum_{k=1}^K \sum_{j=1}^J \mathbf{1}[\entry_k \in
        \binarea_j]\,[\ln\completeness(\entry_k)+\ratepar_j]
    -\sum_{j=1}^J\exp(\ratepar_j)\,
        \int_{\binarea_j} \completeness(\entry)\dd\entry
\end{eqnarray}
where the indicator function $\mathbf{1}[\cdot]$ is one if $\cdot$ is true and
zero otherwise.
Taking the gradient of this function with respect to \ratepars\ and setting it
equal to zero, we find the maximum likelihood result
\begin{eqnarray}\eqlabel{ml-ide}
\exp ({\ratepar_j}^*) &=&
\frac{K_j}{\int_{\binarea_j} \completeness(\entry)\dd\entry}
\end{eqnarray}
where $K_j$ is the number of objects that fall within the bin $j$.
We estimate the uncertainty $\delta\ratepar_j$ on this value by examining the
curvature of the log-likelihood function near the maximum and find
\begin{eqnarray}
\frac{\delta\exp({\ratepar_j}^*)}{\exp({\ratepar_j}^*)}
&=& \frac{1}{\sqrt{K_j}}
\quad.
\end{eqnarray}

In our pathological example from above, the integral of the completeness
function over the bin is $1/2$, giving each sample the expected weight of $2$.
In more realistic cases, where the completeness function varies smoothly, the
inverse-detection-efficiency result will begin to agree with \eq{ml-ide} but
the severity of this bias will be very problem dependent.
Therefore, if you have a dataset with negligible observational uncertainties,
we recommend that you always apply \eq{ml-ide} instead of the standard
inverse-detection-efficiency procedure.
As the uncertainties become more significant, there is no longer an analytic
result and the method derived in this \paper\ is necessary.

\newcommand{\arxiv}[1]{\href{http://arxiv.org/abs/#1}{arXiv:#1}}

\clearpage

\begin{figure}[p]
\begin{center}
\includegraphics[width=\textwidth]{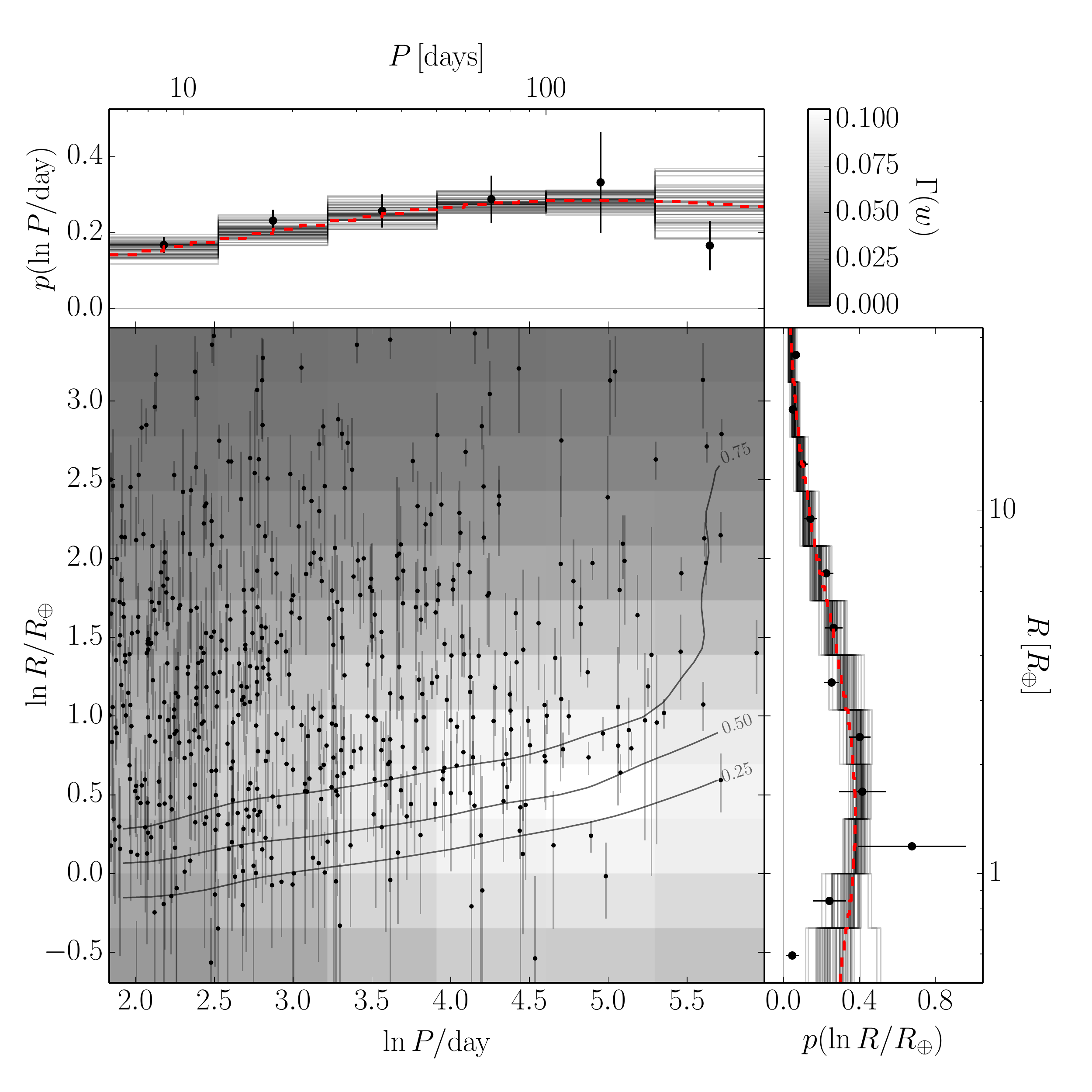}
\end{center}
\caption{%
{\bf Simulated data}.
Inferences about the rate density based on the simulated catalog \modela.
\emph{Center:} the points with error bars show the exoplanet candidates in the
simulated incomplete catalog, the contours show the survey completeness
function (\citealt{petigura}), and the grayscale shows the median posterior
occurrence surface.
\emph{Top and left:} the red dashed line shows the true distribution that was
used to generate the catalog, the points with error bars show the results of
the inverse-detection-efficiency procedure, and the histograms are posterior
samples from the marginalized rate density as inferred by our method.
\figlabel{smooth-results}}
\end{figure}

\begin{figure}[p]
\begin{center}
\includegraphics[width=\textwidth]{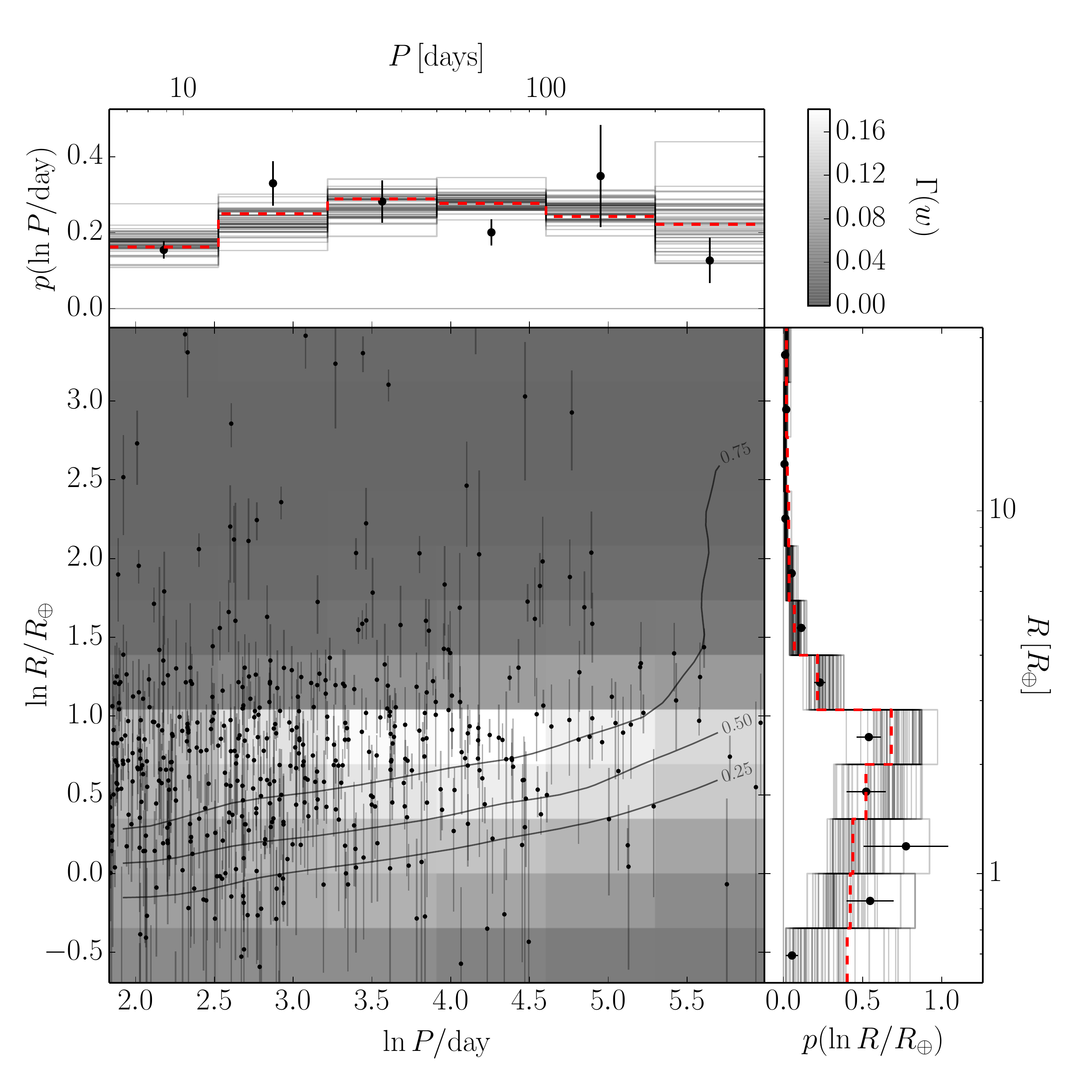}
\end{center}
\caption{%
{\bf Simulated data}.
The same as \fig{smooth-results} for \modelb.
\figlabel{simulation-results}}
\end{figure}

\begin{figure}[p]
\begin{center}
\includegraphics[width=0.8\textwidth]{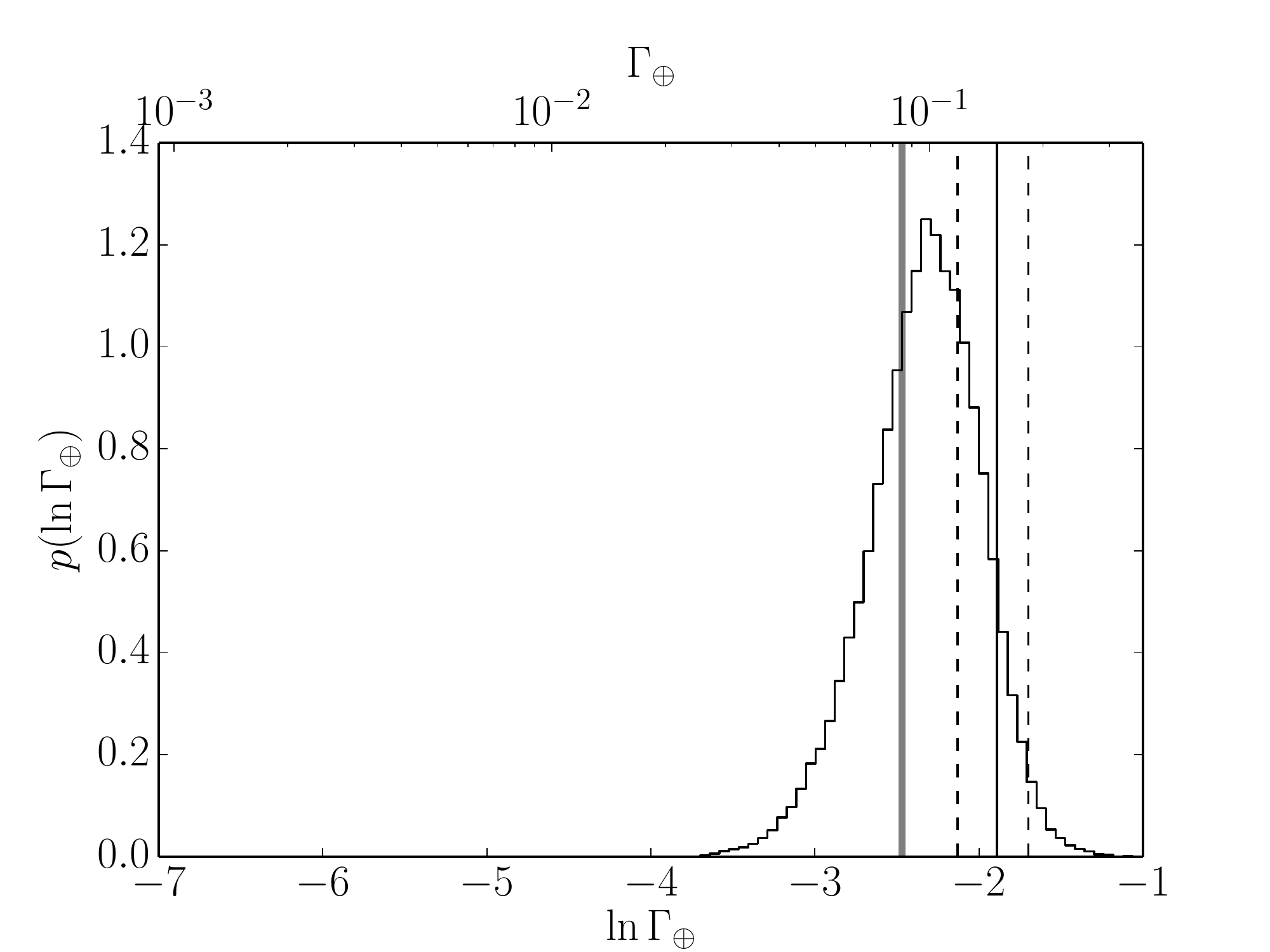}
\end{center}
\caption{%
{\bf Simulated data}.
The extrapolated rate density of Earth analogs \gammaearth\ as inferred by the
different techniques applied to the \modela\ simulation.
Applying the method used by \citet{petigura} gives a constraint indicated by
the vertical black line with error bars shown as dashed lines.
The histogram is the MCMC estimate of our posterior constraint on this rate
density and the true value is indicated as the thick gray vertical line.
\figlabel{smooth-rate}}
\end{figure}

\begin{figure}[p]
\begin{center}
\includegraphics[width=0.8\textwidth]{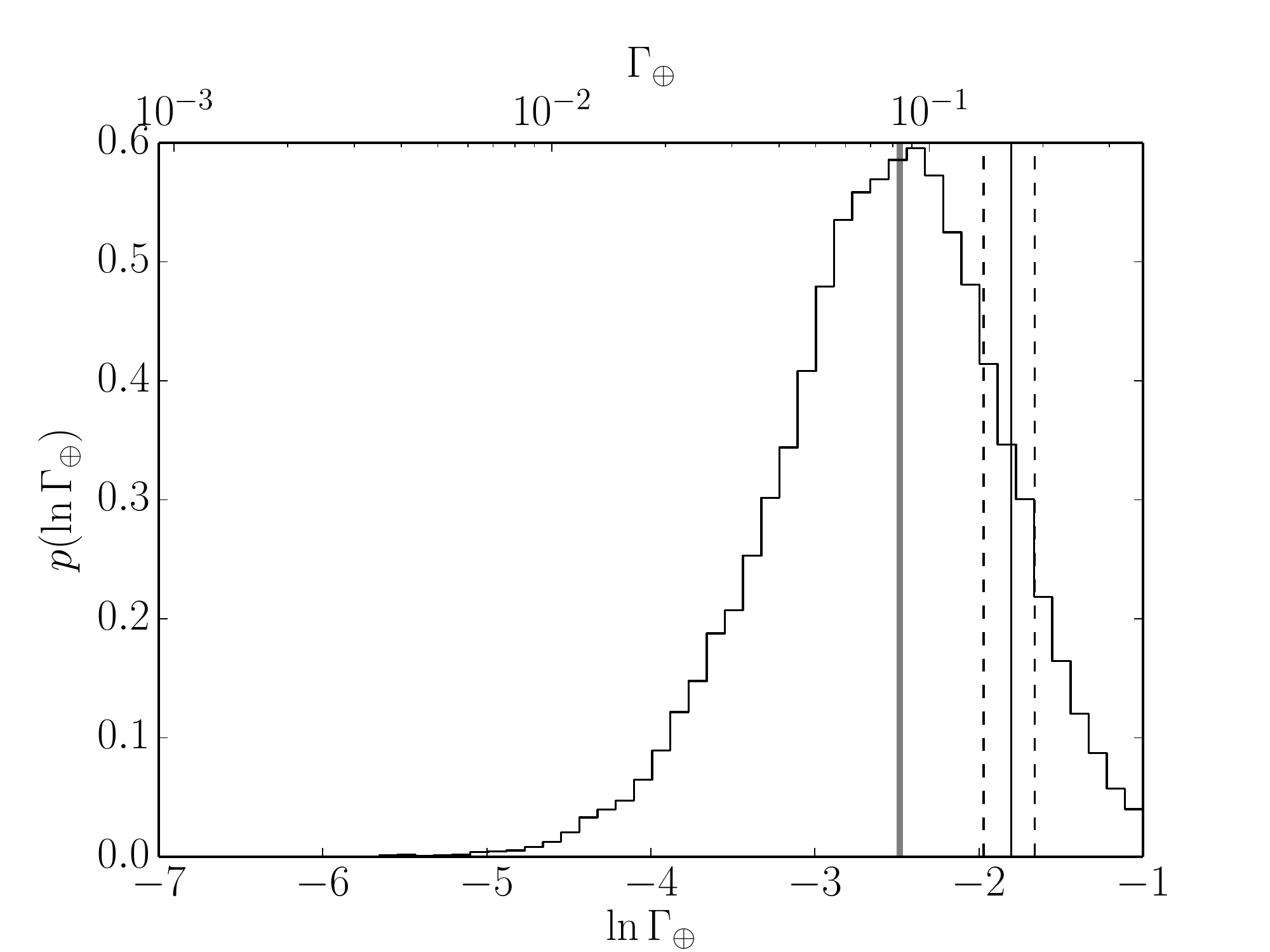}
\end{center}
\caption{%
{\bf Simulated data}.
The same as \fig{smooth-rate} for \modelb.
\figlabel{simulation-rate}}
\end{figure}

\begin{figure}[p]
\begin{center}
\includegraphics[width=\textwidth]{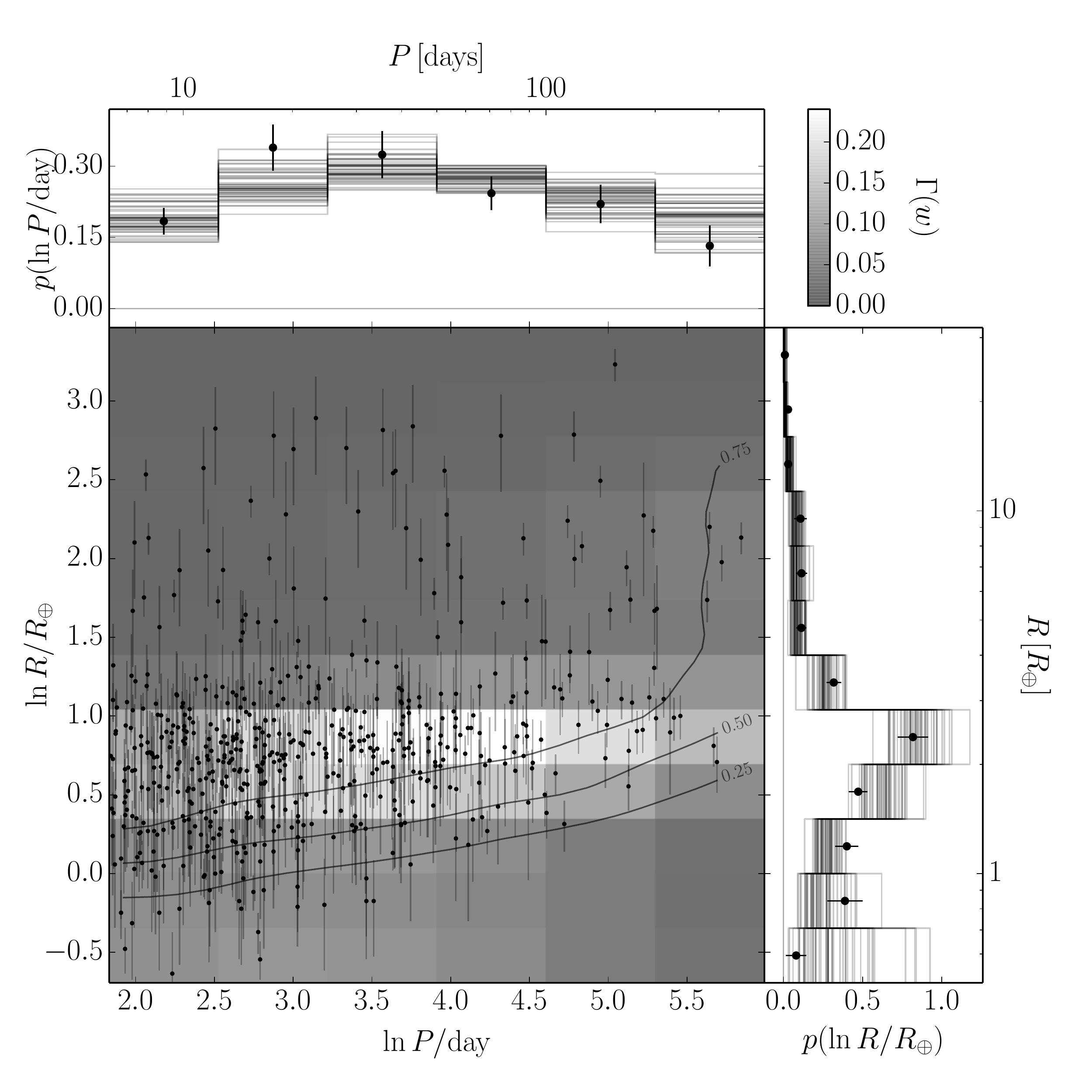}
\end{center}
\caption{%
{\bf Real data}.
The same as \fig{smooth-results} when applied to the observed data from
\citet{petigura}.
\emph{Center:} the points with error bars show the catalog measurements, the
contours show the survey completeness function, and the grayscale shows the
median posterior occurrence surface.
\emph{Top and left:} the points with error bars show the results of the
inverse-detection-efficiency procedure, and the histograms are posterior
samples from the marginalized rate density as inferred by our method.
\figlabel{real-results}}
\end{figure}

\begin{figure}[p]
\begin{center}
\includegraphics{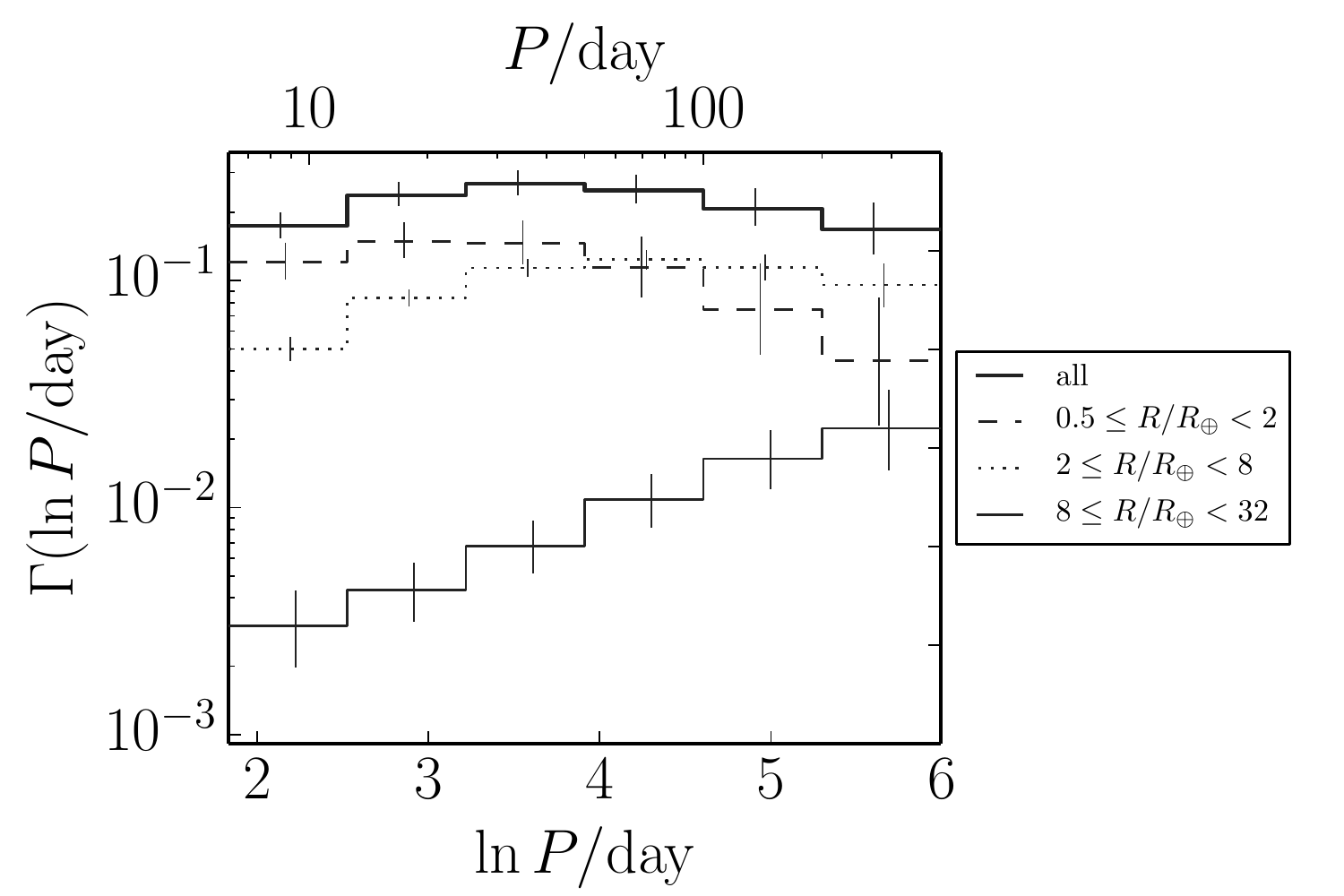}
\end{center}
\caption{%
{\bf Real data}.
The occurrence rate density as a function of logarithmic period integrated
over bins in logarithmic radius.
The lines with error bars show the posterior sample median and 68th
percentile and the line style specifies the radius bin.
The period distribution for the largest planets in the sample
($8 \le R/R_\oplus < 32$) continues to increase (as a function of
$\ln\period$) for all periods while the distribution seems to flatten and
turn over at periods around 50 days.
\figlabel{period}}
\end{figure}

\begin{figure}[p]
\begin{center}
\includegraphics{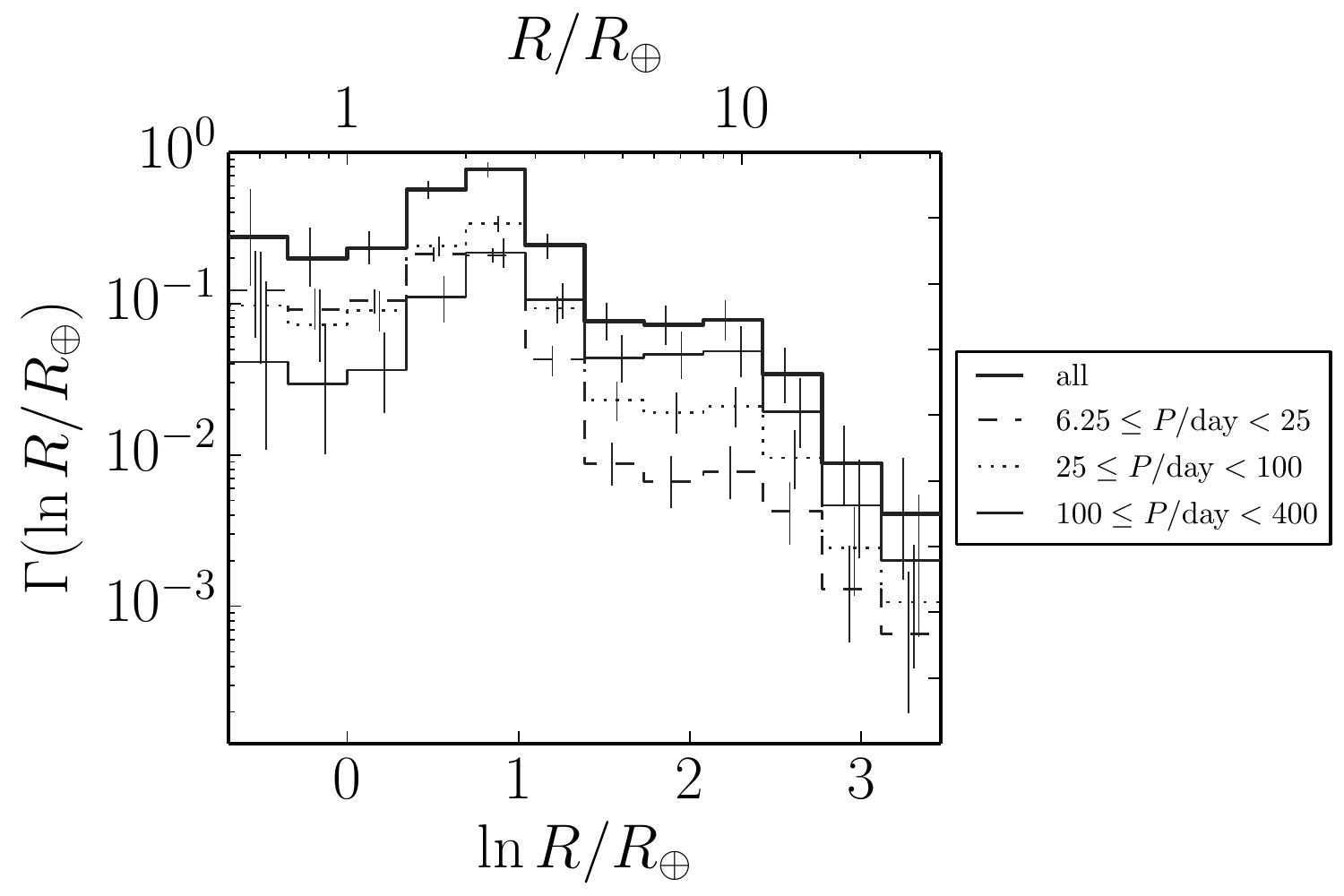}
\end{center}
\caption{%
{\bf Real data}.
The occurrence rate density as a function of logarithmic radius integrated
over bins in logarithmic period.
The lines with error bars show the posterior sample median and 68th
percentile and the line style specifies the period bin.
The distributions in all the period bins are qualitatively consistent and
there are plausibly features near $R\sim3\,R_\oplus$ and $R\sim10\,R_\oplus$.
\figlabel{radius}}
\end{figure}

\begin{figure}[p]
\begin{center}
\includegraphics{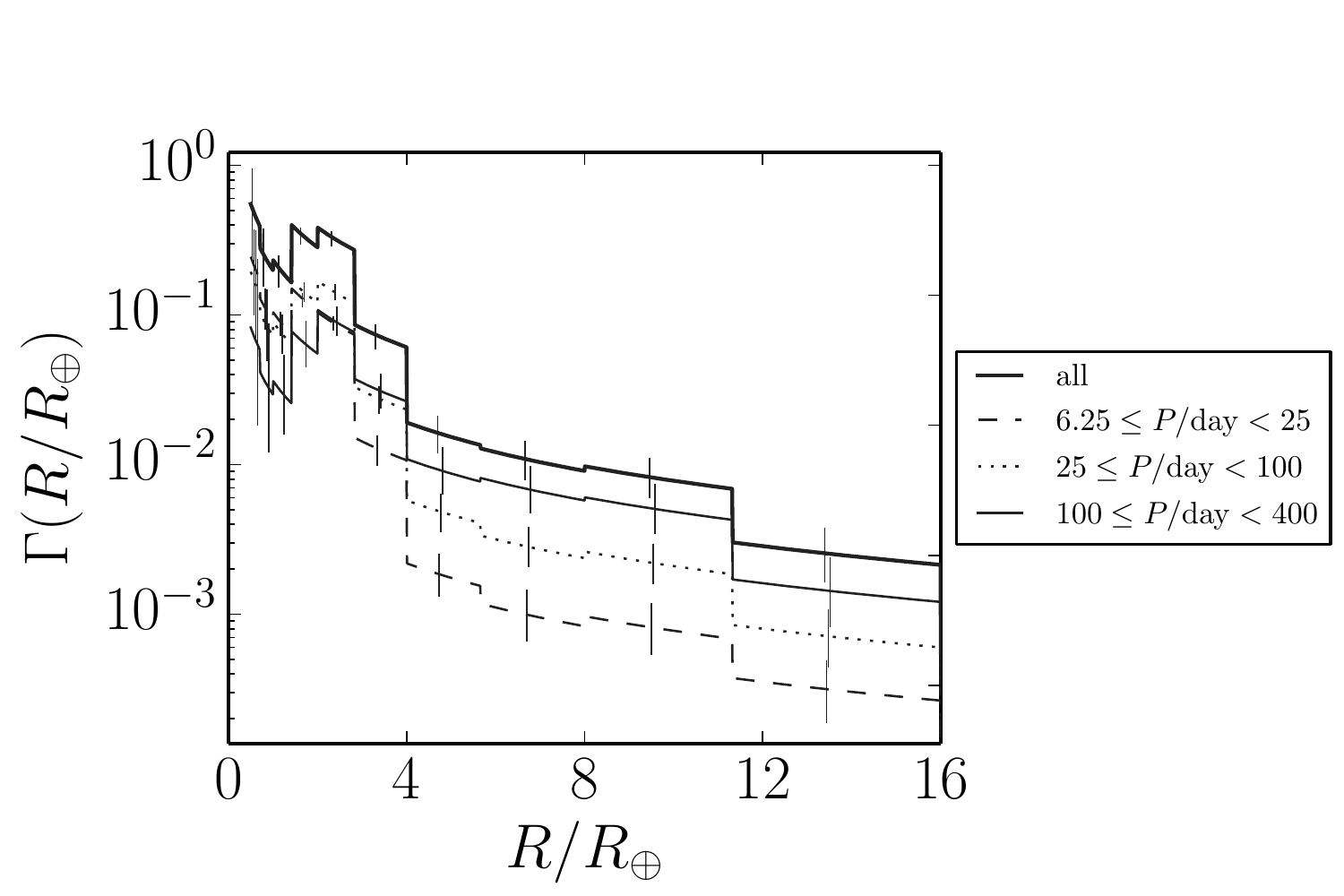}
\end{center}
\caption{%
{\bf Real data}.
The same as \fig{radius} but presented as a density in radius instead of
logarithmic radius.
\figlabel{linear-radius}}
\end{figure}

\begin{figure}[p]
\begin{center}
\includegraphics[width=0.8\textwidth]{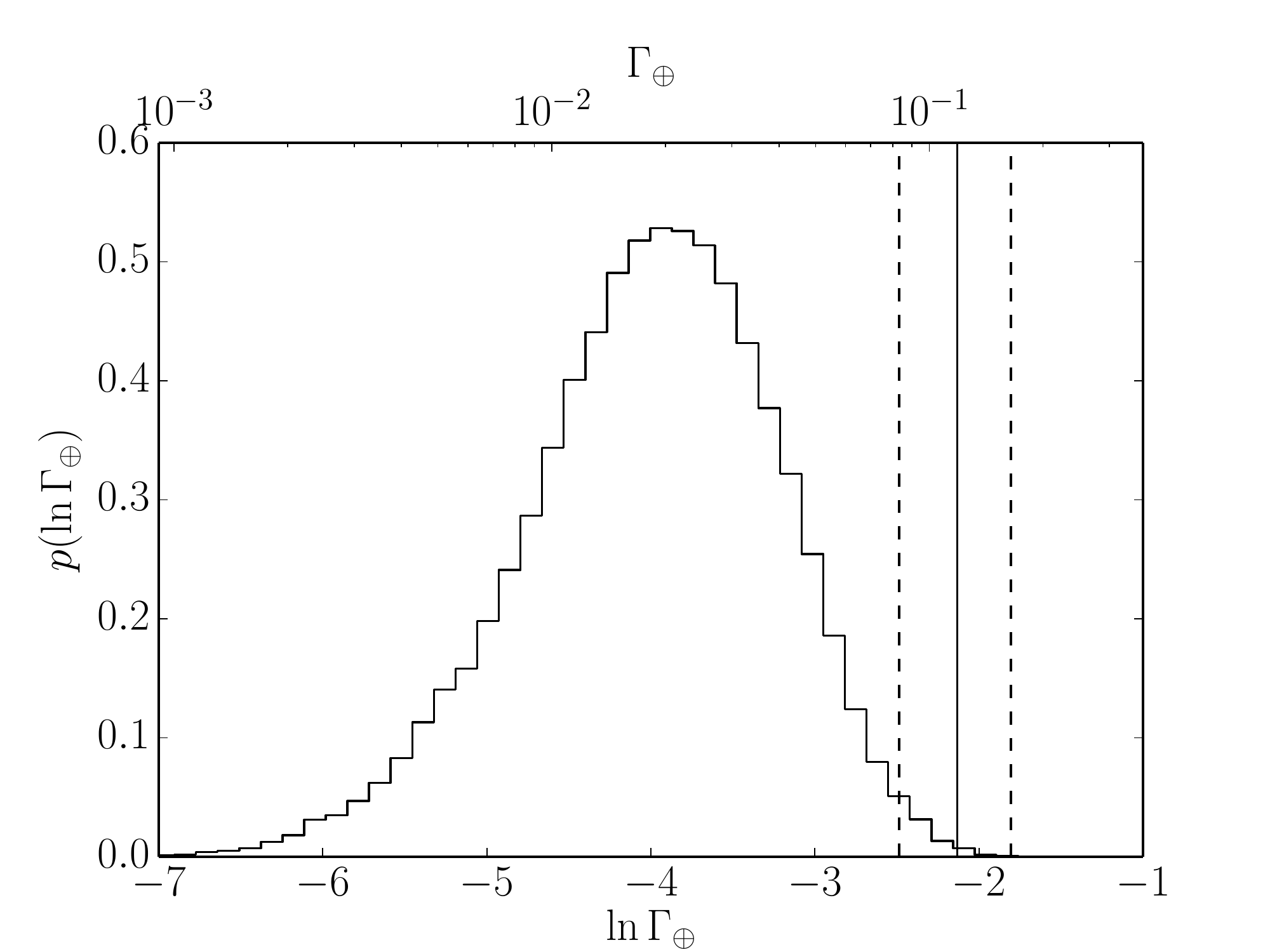}
\end{center}
\caption{%
The extrapolated rate density of Earth analogs \gammaearth\ (the same as
\fig{smooth-rate} but applied to the catalog from \citealt{petigura}).
The histogram is the MCMC estimate of our posterior constraint on this rate
density.
The vertical black line with error bars shown as dashed lines is the result
from \citet{petigura} converted to a rate density by dividing by their bin
volume.
\figlabel{real-rate}}
\end{figure}

\begin{figure}[p]
\begin{center}
\includegraphics[width=\textwidth]{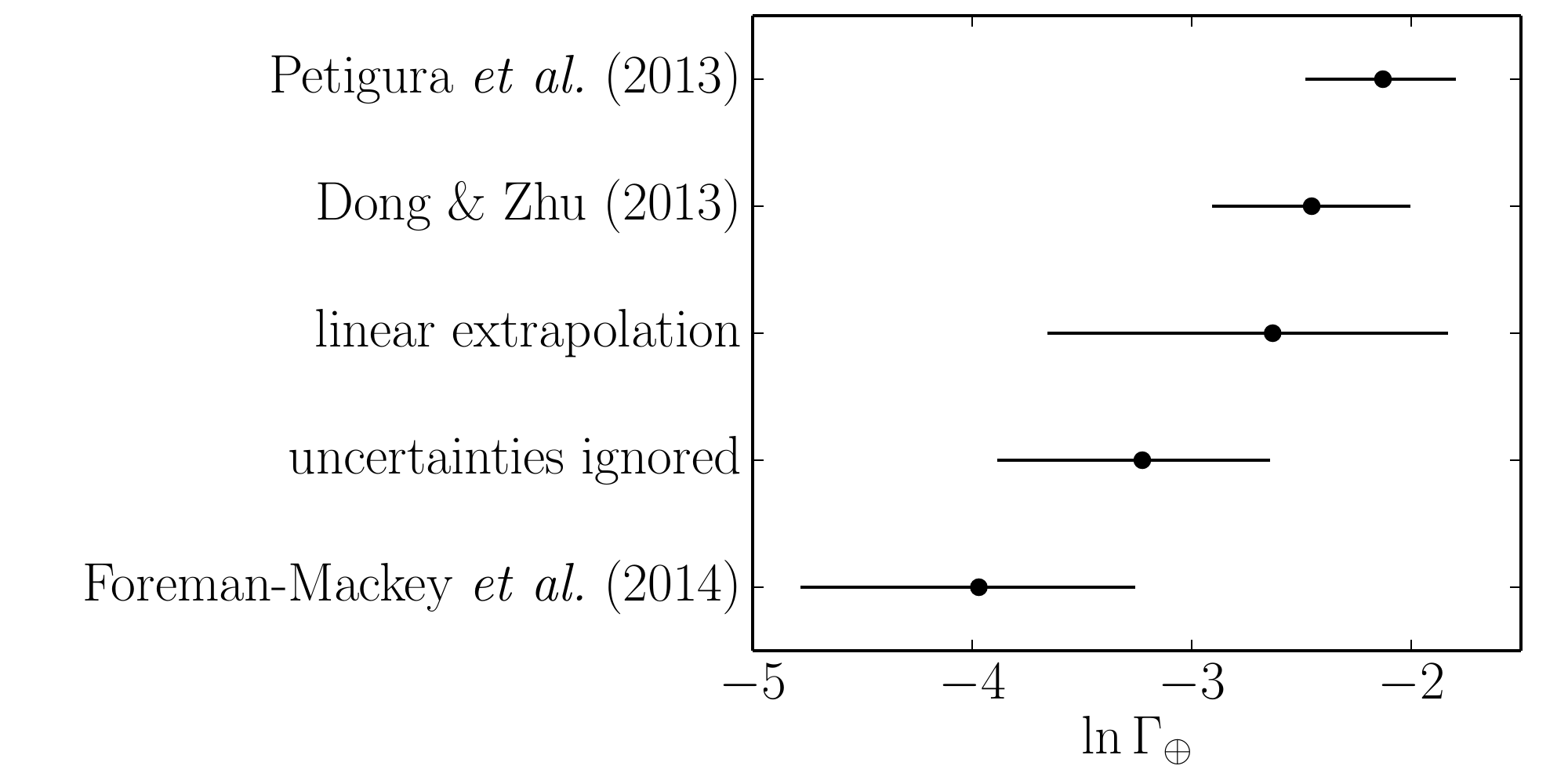}
\end{center}
\caption{%
Comparison of various estimates of \gammaearth.
From the top, the first value is the number published by \citet{petigura} and
converted to consistent units.
The second point shows the value implied by the power law model for the
occurrence rate of $1-2\,R_\oplus$ planets from \citet{dong}.
The point labeled ``linear extrapolation'' is the result of modeling the
distribution of small planets ($1-2\,R_\oplus$) on long periods
($50-400\,\mathrm{days}$) but allowing the period distribution to be
\emph{linear} instead of \emph{uniform}.
The ``uncertainties ignored'' value is given by applying the model
developed in this \paper\ but with the error bars on radius artificially
set to zero.
Finally, the bottom point is the result of our full analysis.
\figlabel{comparison}}
\end{figure}

\end{document}